\documentclass{emulateapj}

\usepackage{graphics}
\usepackage{epsfig}
\usepackage{natbib}
\shorttitle{COS Observations of SN~1987A}
\shortauthors{France et al.}

\begin{document}


\title{$HST$-COS Observations of Hydrogen, Helium, Carbon and Nitrogen Emission from the SN~1987A Reverse Shock\altaffilmark{*}}


\author{
Kevin France\altaffilmark{1}, Richard McCray\altaffilmark{2},
Steven V. Penton\altaffilmark{1}, 
Robert P. Kirshner\altaffilmark{3}, Peter Challis\altaffilmark{3}, 
J. Martin Laming\altaffilmark{4},
Patrice Bouchet\altaffilmark{5},
Roger Chevalier\altaffilmark{6},
Claes Fransson\altaffilmark{7},
Peter M. Garnavich\altaffilmark{8},
Kevin Heng\altaffilmark{9}, 
Josefin Larsson\altaffilmark{7},
Stephen Lawrence\altaffilmark{10},
Peter Lundqvist\altaffilmark{7},
Nino Panagia\altaffilmark{11,12,13},
Chun S. J. Pun\altaffilmark{14},
Nathan Smith\altaffilmark{15},
Jesper Sollerman\altaffilmark{7},
George Sonneborn\altaffilmark{16},
Ben Sugerman\altaffilmark{17},
J. Craig Wheeler\altaffilmark{18}
}
\altaffiltext{*}{Based on observations made with the NASA/ESA $Hubble$~S$pace$~$Telescope$, obtained from the data archive at the Space Telescope Science Institute. STScI is operated by the Association of Universities for Research in Astronomy, Inc. under NASA contract NAS 5-26555.}



\begin{abstract}

We present the most sensitive ultraviolet observations of Supernova 1987A to date.  
Imaging spectroscopy from the {\it Hubble Space Telescope}-Cosmic Origins Spectrograph shows many narrow ($\Delta$$v$~$\sim$~300 km~s$^{-1}$) emission lines from the circumstellar ring, broad ($\Delta$$v$~$\sim$~10~--~20~$\times$~10$^{3}$ km~s$^{-1}$) emission lines from the reverse shock, and ultraviolet continuum emission.  The high signal-to-noise ($>$ 40 per resolution element) broad Ly$\alpha$ emission is excited by soft X-ray and EUV heating of mostly neutral gas in the circumstellar ring and outer supernova debris. The ultraviolet continuum at $\lambda$~$>$~1350~\AA\ can be explained by \ion{H}{1} 2-photon (2$s$~$^{2}S_{1/2}$~--~1$s$~$^{2}S_{1/2}$) emission from the same region. We confirm our earlier, tentative detection of \ion{N}{5} $\lambda$1240 emission from the reverse shock and present the first detections of broad \ion{He}{2} $\lambda$1640, \ion{C}{4} $\lambda$1550, and \ion{N}{4}] $\lambda$1486 emission lines from the reverse shock. The helium abundance in the high-velocity material is He/H = 0.14~$\pm$~0.06.  The \ion{N}{5}/H$\alpha$ line ratio requires partial ion-electron equilibration ($T_{e}$/$T_{p}$~$\approx$~0.14~--~0.35).  We find that the N/C abundance ratio in the gas crossing the reverse shock is significantly higher than that in the circumstellar ring, a result that may be attributed to chemical stratification in the outer envelope of the supernova progenitor.  The N/C abundance ratio may have been stratified prior to the ring expulsion, or this result may indicate continued CNO processing in the progenitor subsequent to the expulsion of the circumstellar ring.  

\end{abstract}



\clearpage


\section{Introduction}
Borkowski, Blondin, \& McCray (1997) 
predicted that the spectrum of SN~1987A should display very broad 
($\Delta$$v$~$\sim$~$\pm$~12,000 km~s$^{-1}$) 
emission lines of Ly$\alpha$, H$\alpha$, \ion{N}{5}~$\lambda$1240, and \ion{He}{2}~$\lambda$1640, produced where the freely expanding supernova debris crosses a reverse shock located inside the equatorial circumstellar ring. In September 1997, using the {\it Hubble Space Telescope}-Space Telescope Imaging Spectrograph ($HST$-STIS), \citet{sonneborn98} detected broad Ly$\alpha$ emission, and~\citet{michael98} showed how observations of this emission can be used to map the shape of the reverse shock and the flux of \ion{H}{1} atoms crossing it.  Michael et al (2003) and~\citet{heng06} analyzed subsequent (February 1999~--~ October 2002) STIS observations of both Ly$\alpha$ and H$\alpha$ to map the increasing flux of \ion{H}{1} atoms across the reverse shock.\nocite{sonneborn98,michael03,heng06,fransson11} 

The evolution of the broad H$\alpha$ emission can also be tracked with ground-based telescopes, notably in February 2005 with the Magellan telescope~\citep{smith05} and from December 2000 to January 2009 with the Very Large Telescope (Fransson et al 2011).  Smith et al. pointed out that the H$\alpha$ emission from the reverse shock could be suppressed due to preionization of hydrogen in the supernova debris by soft X-rays and EUV radiation from the rapidly brightening shock interaction with the inner circumstellar ring. Extrapolating the X-ray light curve, they predicted that this preionization would cause the broad H$\alpha$ emission to vanish by 2012.  However, since 2005, the X-ray light curve has leveled off~\citep{park11}, so the preionization effect should be less than that predicted by Smith et al. 

\citet{france10c} analyzed the most recent (31 January 2010) STIS observations of SN~1987A. They pointed out that the observed ratio of Ly$\alpha$/H$\alpha$ photon fluxes was~$\gtrsim$~30 at large, negative velocities, much greater than the value~$\approx$~5 that would be expected for hydrogen atoms excited as they cross the reverse shock. Moreover, the spatially resolved Ly$\alpha$ line profile differed dramatically from that of H$\alpha$.  France et al. proposed that the broad Ly$\alpha$ emission is dominated by Ly$\alpha$ emission from the nearly stationary equatorial ring that has been resonantly scattered by hydrogen atoms in the expanding supernova debris.  France et al. also noticed a faint glow at wavelengths ranging from about 1260~--~1290 \AA, which they attributed to the \ion{N}{5}~$\lambda$1240 emission predicted by Borkowski et al.  They suggested that a critical test of that hypothesis would be detection of broad emission from \ion{C}{4}$\lambda$1550.

Here we describe far-ultraviolet (1140~--~1780 \AA) spectra of SN~1987A obtained with the Cosmic Origins Spectrograph (COS) that was installed on $HST$ in May 2009.  With COS, we have measured the profiles of Ly$\alpha$ and \ion{N}{5}~$\lambda$1240 with signal-to-noise ratios (S/N) far superior to the previous STIS spectra.  We have also detected the broad \ion{He}{2}~$\lambda$1640 emission predicted by~\citet{borkowski97} and the broad \ion{C}{4}~$\lambda$1550 emission predicted by France et al (2010).  We find that while the He/H abundance ratio in the outer ejecta is consistent with that of the circumstellar ring, the N/C ratio is enhanced by more than a factor of two over the ring abundances.     

\begin{figure}
 \begin{center}
 \epsfig{figure=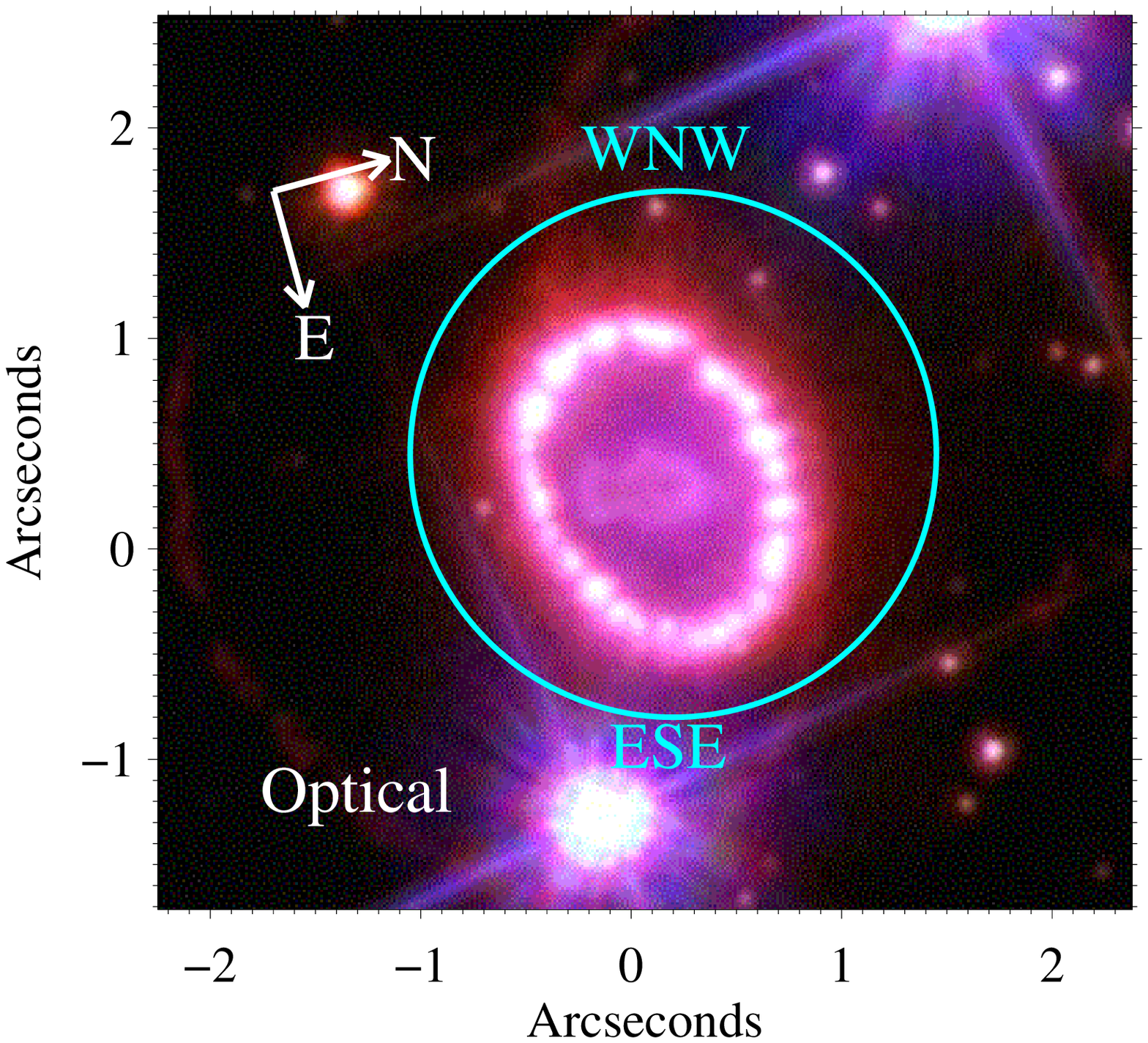,width=3.5in,angle=90}
 \epsfig{figure=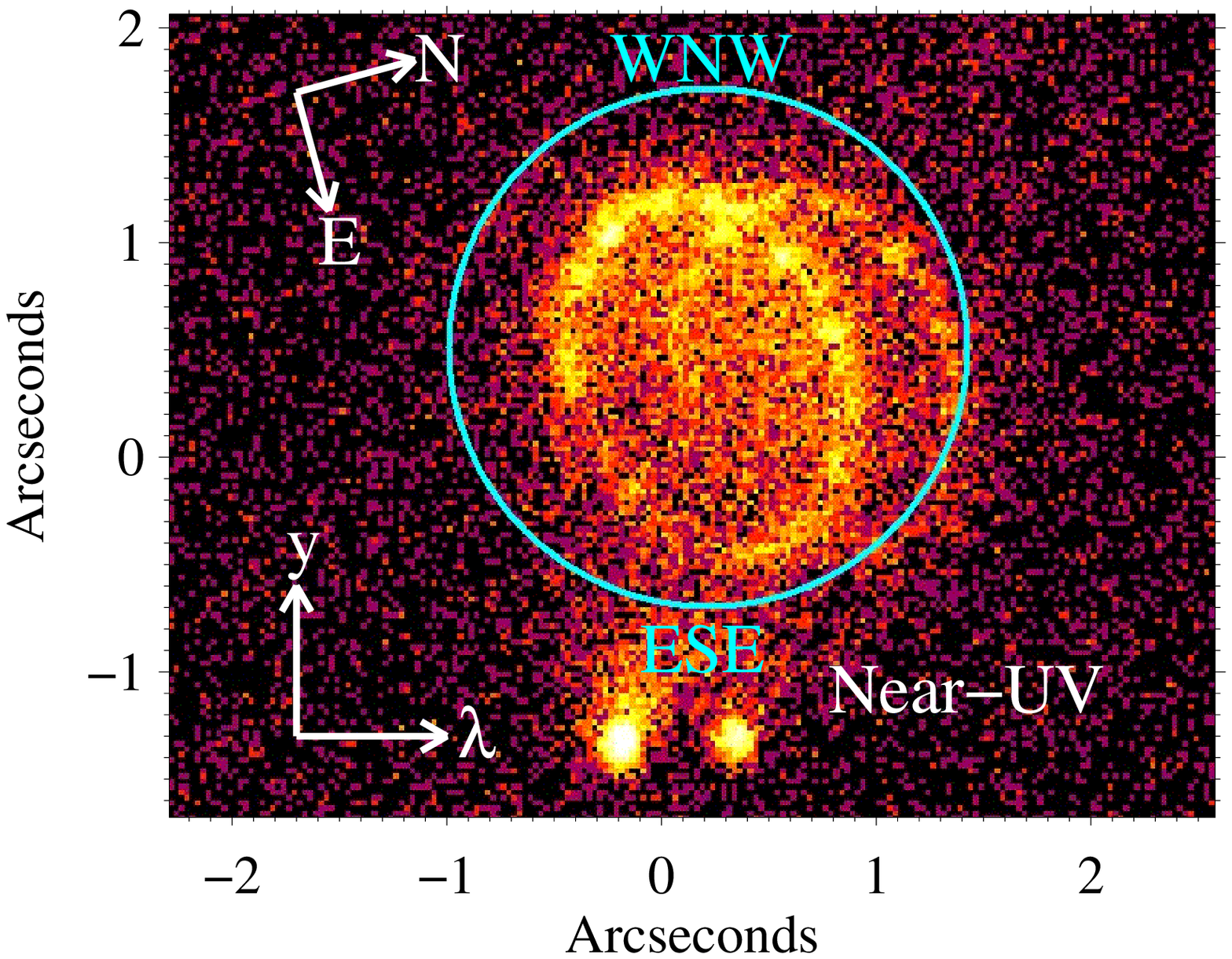,width=3.5in,angle=90}
 \caption{\label{cosdata} ($top$) $HST$-ACS $BVR$ image with a representative overlay of the 2.5\arcsec\ diameter COS aperture (cyan).  
($bottom$)
$HST$-COS near-UV pointing verification image for the G130M (2011 February 11) visit.  Due to bright limits, this observation was made using the MIRRORB imaging configuration, which creates the image doubling.    Because COS is a slitless spectrograph, light from outside the nominal 2.5\arcsec\ can enter the system.  Star 3, separated from the center of the circumstellar ring by ~$\approx$~1.7\arcsec\ is seen at the bottom.  In order to match the spectral image presented in Figure 2, we display the coordinates of this image with east to the right of north, opposite from the standard convention.  The spectroscopic dispersion axis is labeled ``$\lambda$'' and the cross-dispersion axis is labeled ``y''.  
  }
 \end{center}
 \end{figure}

\section{$HST$-COS Observations, Imaging Spectroscopy, and Data Reduction}
SN~1987A was observed with the medium resolution far-UV modes of $HST$-COS (G130M and G160M) on 2011 February 11 and March 14 for a total of 7 spacecraft orbits (18555 s; Table 1) as part of the Supernova 1987A INTensive Study (SAINTS - GO12241; PI - R. Kirshner).  A description of the COS instrument and on-orbit performance characteristics can be found in~\citet{osterman11}.  All observations were approximately centered on the SN~1987A circumstellar ring (R.A. = 05$^{\mathrm h}$ 35$^{\mathrm m}$ 28.07$^{\mathrm s}$, Dec. = -69\arcdeg 16\arcmin 10.8\arcsec ; J2000) and COS performed an offset imaging target acquisition from a reference star $\approx$ 8\arcsec\ away. A NUV image was obtained with the MIRRORB imaging mode after the slew to the science pointing in order to verify that the circumstellar ring was in the primary science aperture (PSA).  Figure 1 ($top$) shows the COS aperture location on an optical image of SN~1987A,  The MIRRORB configuration introduces optical distortions into the image, but a first-order analysis indicates that the entire circumstellar ring was inside the 2.5\arcsec\ diameter PSA (Figure 1, $bottom$).

The COS position angle was chosen to maximize the spatial separation on the COS detector between Star 3 and the supernova emission we aim to study.
The G130M observations were made at a position angle of $\approx$~$-$75\arcdeg, while the G160M observations were acquired with a position angle of $\approx$~$-$45\arcdeg.    Light from objects outside the nominal 1.25\arcsec\ COS aperture radius can be recorded with the science spectrum in crowded fields.  Star 3 is a far-UV-bright Be star~\citep{gilmozzi87,walborn93} separated from the center of the SN~1987A ring by ~$\approx$~1.7\arcsec.  Stellar contamination could compromise supernova data quality if the two objects overlapped on the spectroscopic detector. Thus, we used the imaging spectroscopic capability of COS to keep SN~1987A  
centered while placing Star 3 at the bottom of the  microchannel plate (MCP) detector.  

We complemented this observing strategy with post-processing techniques and custom spectral extractions to maximize the spatial resolution of the instrument in the cross-dispersion direction.  Due to long-term exposure of the COS MCP to geocoronal Ly$\alpha$ and hot-star spectra, the number of electrons generated by each incident photon at a given detector location has been decreasing.  This is manifest as lower pulse heights observed by the detector electronics\footnote{We refer the reader to the cycle 19 COS Instrument Handbook for more details: {\tt http://www.stsci.edu/hst/cos/documents/handbooks/current/cos\_cover.html}}.  The exact form of the pulse-height evolution is a complicated function of the dispersion and cross-dispersion arrival location of the incident photon, but to first order we can approximate this as a pulse-height dependent mislocation of the photon $y$-location.  The approximate form of this mislocation is -0.47 $y$ pixels per pulse height bin drop.  We have reprocessed the COS observations of SN~1987A, including a correction for the cross-dispersion misplacement, producing more accurate photon arrival  positions.  
This additional data reduction step improves the far-UV angular resolution of the instrument by 20~--~30\% across the bandpass.  

\begin{figure}[b]
 \begin{center}
 \epsfig{figure=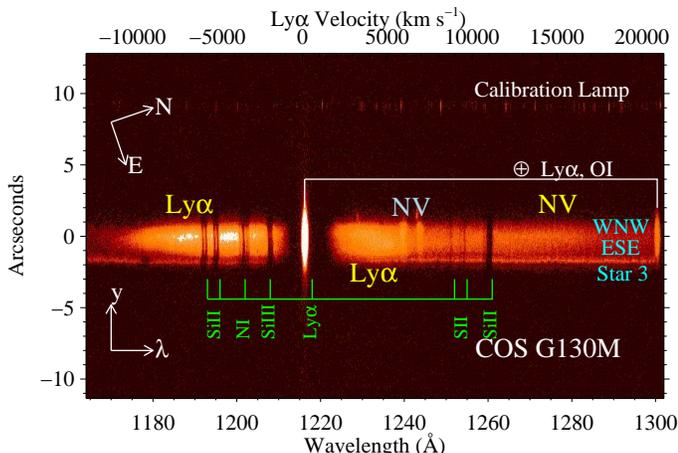,width=2.75in,angle=90}
 \caption{\label{cosdata} Two-dimensional spectrogram of the G130M segment B data.  Reverse shock emission is labeled in yellow, hotspot emission is labeled in blue, and interstellar absorption features are marked in green.   
  }
 \end{center}
 \end{figure}

 \begin{figure}[t]
 \begin{center}
 \epsfig{figure=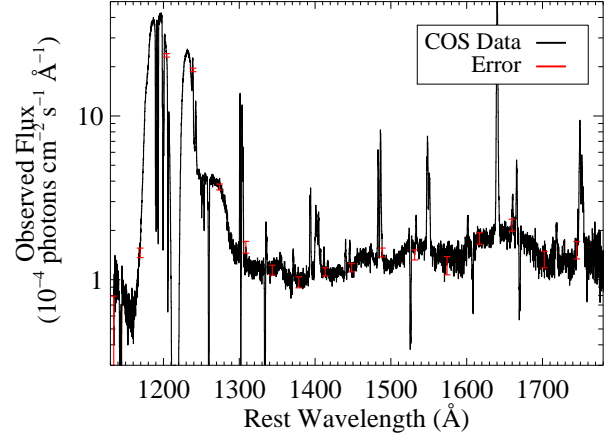,width=2.6in,angle=90}
 \caption{\label{cosdata} Full far-UV spectrum of the combined WNW and ESE regions.  Broad emission from the SN~1987A reverse shock, narrow emission lines attributable to circumstellar hotspots, as well as underlying continuum emission are observed. Representative 1-$\sigma$ error bars 
(a combination of photon statistics and flux calibration uncertainties) are shown in red.  
  }
 \end{center}
 \end{figure}

Star 3 can be seen at the bottom of the two-dimensional spectrogram shown in Figure 2.  Star 3 is well isolated at the bottom of the detector.  The angular resolution in the reprocessed two-dimensional data is $\approx$~0.8\arcsec.
This imaging capability allows us to make custom spectral extractions at three spatially resolved locations: the ``WNW'' region at the top of the detector (approximately centered on the hotspots at: R.A. = 05$^{\mathrm h}$ 35$^{\mathrm m}$ 27.98$^{\mathrm s}$, Dec. = -69\arcdeg 16\arcmin 10.5\arcsec ; J2000), the "ESE" region towards Star 3 (approximately the telescope pointing coordinates), and a separate extraction of Star 3 itself.   
Star 3 contributes negligible flux to the one-dimensional WNW spectrum and less than 15\% of the flux in the ESE extraction. 
The spatially resolved spectra will be analyzed in a future work, and we focus on the combined WNW~+~ESE spectra here.

The custom data extractions were then reprocessed with the COS calibration pipeline, CALCOS v2.13.6, and combined with the custom IDL coaddition procedure described by~\citet{danforth10} and~\citet{shull10}. 
The coaddition routine interpolates all detector segments and grating settings onto a common wavelength grid, and makes a correction for the detector QE-enhancement grid.  No correction for the detector hex pattern is performed.  Data were obtained in four central wavelength settings in each far-UV grating mode ($\lambda$1291, 1300, 1309, and 1318 with G130M and $\lambda$1577, 1589, 1600, and 1611 with G160M) at the default focal-plane split position. The instrumental configurations are summarized in Table 1.  Observations at multiple wavelength settings provide continuous spectral coverage over the 1136~--~1782~\AA\ bandpass and minimize the residual fixed pattern noise from the detector grid wires and the MCP pores.  The point source resolving power of the medium resolution COS far-UV modes is $R$~$\equiv$~$\Delta\lambda$/$\lambda$~$\approx$~18,000 ($\Delta$$v$~=~17 km s$^{-1}$); however, the filled-aperture resolving power is $R$~$\sim$~1500 ($\Delta$$v$~$\sim$~200 km s$^{-1}$; France et al. 2009).\nocite{france09}  Multiple point sources within the aperture (e.g., the multiple hotspots along the SN~1987A circumstellar ring) produce spectral resolution that is similar to the extended source response.  The point source flux calibration of COS is accurate to better than~$\approx$~3\%; however, the extended source and custom processing limit the absolute flux calibration of the SN~1987A far-UV spectroscopic data to $\sim$~10\%.

\section{Data Analysis}

\subsection{Emission Spectrum and Interstellar Corrections}

The extracted emission spectrum of SN~1987A is shown in Figure 3.  The spectra presented here are shown in photon units as these are the natural units for line flux comparisons.  
 There is a wealth of narrow and broad emission and absorption features in the spectrum as well as a strong underlying continuum.  The ``narrow'' (FWHM~$\approx$~200~--~300 km s$^{-1}$) resolved emission features are attributed to the circumstellar ring.  The circumstellar ring spectrum is dominated by the hotspots which originate at shock interfaces where the forward blast wave encounters regions of high density in the circumstellar ring (Lawrence et al. 2000).  A discussion of the ring-ejecta interaction is given by Larsson et al. (2011).~\nocite{lawrence00,larsson11}
Most of the circumstellar ring emission lines have been observed in previous $HST$-STIS (e.g., Pun et al. 2002) and $IUE$ (Lundqvist \& Fransson 1996) observations, including \ion{N}{5} $\lambda$1239, 1243, \ion{C}{2} $\lambda$1334, 1335, \ion{Si}{4} $\lambda$1394, 1403, the \ion{O}{4}] $\lambda$1400 multiplet, 
\ion{N}{4}] $\lambda$1483, 1486, \ion{C}{4} $\lambda$1548, 1550, [\ion{Ne}{4}] $\lambda$1602, \ion{He}{2} $\lambda$1640, \ion{O}{3}] $\lambda$1661, 1666, and the \ion{N}{3}] $\lambda$1750 multiplet.  Our deep COS observations enable first detections of \ion{O}{5} $\lambda$1371 and \ion{Si}{2} $\lambda$1526, 1533, although the ground state transition of \ion{Si}{2} is blocked by the interstellar medium.  We also tentatively detect the coronal [\ion{Si}{8}] doublet $\lambda$1441, 1446.  
These features will be discussed in detail in a future work, though we will make use of the circumstellar ring line fluxes where relevant.  

\begin{deluxetable*}{ccccc}
\tabletypesize{\small}
\tablecaption{SN~1987A COS observing log. \label{cos_obs}}
\tablewidth{0pt}
\tablehead{
\colhead{Date}  & \colhead{COS Mode} & \colhead{Central Wavelengths}   & \colhead{FP-POS} 
& \colhead{T$_{exp}$ (s)} 
}
\startdata	
2011 February 11	  &  	G130M 	& 	1291,1300,1309,1218 	& 3 &  	 7884	 \\
2011 March 14	  &  	G160M 	& 	1577,1589,1600,1611 	& 3 &  	 10671     
 \enddata

\end{deluxetable*}

 \begin{figure}
 \begin{center}
 \epsfig{figure=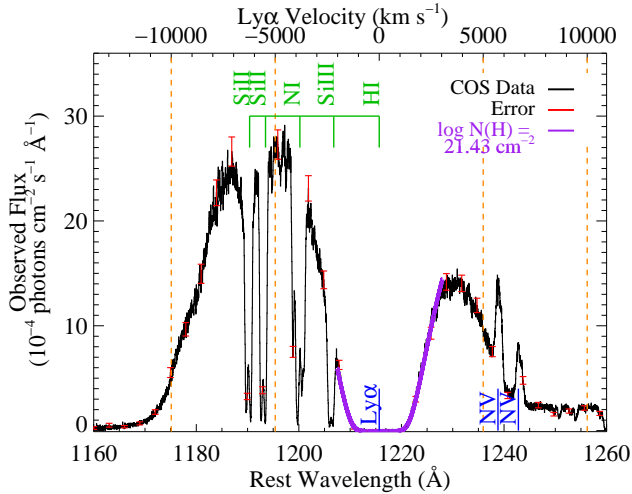,width=2.6in,angle=90}
 \caption{\label{cosdata}   RS Ly$\alpha$ and hotspot \ion{N}{5} in the WNW extraction.  Hotspot emission lines are marked in blue and interstellar absorption components are labeled in green.  The purple line shows a fit to the Ly$\alpha$ line core for log$_{10}$(N(H))~=~21.43 ($\pm$~0.02) cm$^{-2}$.
Dashed orange lines are plotted as velocity references.  Representative 1-$\sigma$ error bars (a combination of photon statistics and flux calibration uncertainties) are shown in red.  
  }
 \end{center}
 \end{figure}

The broad emission features observed in the spectrum of SN~1987A arise as atoms and ions cross the reverse shock front~\citep{michael98}, located just interior to the circumstellar ring.  \ion{H}{1} Ly$\alpha$ (and the corresponding H$\alpha$) emission from the reverse shock has been well studied~\citep{michael03,smith05,heng06,france10c}, and we display the broad ($\Delta$$v$~$\sim$~$-$13,000~--~+7,000 km s$^{-1}$) Ly$\alpha$ emission seen in the COS observations in Figures 4 and 5.  
The second broad feature to the red of Ly$\alpha$ in Figure 3 is consistent with the putative \ion{N}{5} reverse shock emission described by~\citet{france10c} and has a velocity distribution and total flux that are approximately consistent with those predicted by Borkowski et al. (1997); see also \S4.3.1.  In light of these results and the large relative offset from the observed Ly$\alpha$ profile, we claim that reverse shock \ion{N}{5} emission is unambiguously detected in our data (Figure 5).  While
at earlier times, low-ionization UV line emission from SN~1987A has been attributed to the debris in the core~\citep{jerkstrand11}, the core is enshrouded in dust~\citep{matsuura11} and is mostly opaque to UV radiation.
Therefore, the reverse shock is  the most likely formation site for high-velocity UV line photons observed towards SN~1987A.  
Following similar line-identification arguments as those for \ion{N}{5}, we also detect reverse shock emission from \ion{N}{4}] $\lambda$1486, \ion{C}{4} $\lambda$1550, and \ion{He}{2} $\lambda$1640 for the first time.  These features are marked in Figure 6.  

Figure 4 also shows an example of the broad and narrow interstellar absorption features imposed on the spectrum of SN~1987A.  This figure shows Galactic and Magellanic Cloud absorption from \ion{Si}{2} $\lambda$1190, 1193, the \ion{N}{1} $\lambda$1200 multiplet, \ion{Si}{3} $\lambda$1206, and \ion{H}{1} Ly$\alpha$.  
Other interstellar absorbers include \ion{S}{2} $\lambda$1250, 1253, \ion{Si}{2} $\lambda$1260, \ion{C}{2} $\lambda$1334, \ion{Si}{2} $\lambda$1526, \ion{Fe}{2} $\lambda$1608, and \ion{Al}{2} $\lambda$1670.   We fit the interstellar neutral hydrogen absorption in the WNW spectrum as this provides the best measure of the foreground absorption, finding log$_{10}$(N(H))~=~21.43~$\pm$~0.02 cm$^{-2}$.  For subsequent analysis of the Ly$\alpha$ and \ion{N}{5} emission, we divide the data by a model of this emission (shown in purple in Figure 4).  This procedure renders unusable the inner $\pm$~6~\AA\ of the Ly$\alpha$ line profile, which we remove from the data.  

There are several estimates of the interstellar reddening towards the region of the LMC in which SN~1987A resides~\citep{walker90,fitzpatrick90,scuderi96,michael03}.  We adopt a Milky Way ($R_{V}$~=~3.1) extinction curve~\citep{ccm} with $E(B - V)$~=~0.19 as this curve approximates the average extinction correction favored by the studies cited above.  This curve also closely approximates the far-UV properties of the average LMC curve presented by~\citet{gordon03} for $A_{V}$~=~0.6.  Assuming $E(B - V)$~=~0.17~\citep{michael03}, would lower the fluxes derived in subsequent sections by ~$\sim$~15\%.   \citet{scuderi96} present a comprehensive study of the extinction towards Star 2, and the use of this curve would result in fluxes $\sim$~20~--~30\% larger than those derived below.  None of the possible extinction curves significantly alters the shape of the continuum (\S4.1.1) or the far-UV line ratios (\S4.3.3).  

The bright continuum ($F_{\lambda}$(observed)~$\sim$~1~$\times$~10$^{-4}$ photons cm$^{-2}$ s$^{-1}$ \AA$^{-1}$) underlying the hotspot and reverse shock emission was suggested in coadded STIS G140L spectra from 2010, cospatial with the circumstellar ring. However, the significance of the detection was low.   We propose that this continuum is mostly \ion{H}{1} 2-photon emission (\S4.1.1).

\subsection{Binned Spectrum for Broad Line \& Continuum Analysis } 

Narrow emission lines, as well as the narrow interstellar absorption components, complicate measurements of the continuum and broad emission from the reverse shock.  In order to remove confusion from these narrow 
 lines during the analysis of the broad spectral features, we created binned spectra of the WNW and ENE extractions.   These binned spectra were centered on hand-chosen regions of the data at 1~--~5~\AA\ intervals and were free of narrow features.  The flux and error of these binned data points were taken to be the average flux and standard deviation of a 0.8~\AA\ region centered on the selected wavelength.  These binned data are shown as diamonds overplotted on the spectra in Figures 5~--~8.  

We quantify the total reverse shock flux by integrating the binned spectra over wavelength intervals corresponding to the velocity ranges of interest for a given feature.  The velocity ranges were chosen to cover the maximum extent of a given line without significantly overlapping with other reverse shock emission features.  The inner approximately 3000 km s$^{-1}$ were not included to avoid strong hotspot emission lines located at $\approx$~0 km s$^{-1}$.   In most cases the hot spot lines are doublets, therefore a velocity interval larger than the nominal 300 km s$^{-1}$ narrow line-width was excluded.  
We were unable to measure two reverse shock components due to velocity blending.  The blue component of the \ion{N}{5} line is lost under the Ly$\alpha$ line, and the weak red wing of the \ion{N}{4}] emission is overwhelmed by the stronger blue \ion{C}{4} emission.  The velocity and wavelength intervals are given in Table 2 with the summed line fluxes.   

 \begin{figure}
 \begin{center}
 \epsfig{figure=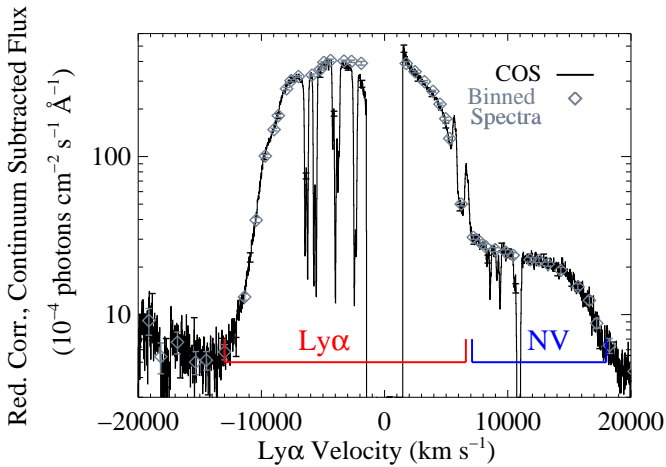,width=2.6in,angle=90}
 \caption{\label{cosdata}   Velocity profile of the ISM corrected (dust and neutral hydrogen), continuum subtracted RS Ly$\alpha$ and \ion{N}{5} in the combined WNW + ESE spectrum.  The binned spectrum described in \S3.2 are shown overplotted in gray.  The Ly$\alpha$ emission extends from $-$13000~--~+7000 km s$^{-1}$ and emission to the red of Ly$\alpha$ is attributed to \ion{N}{5}.   
  }
 \end{center}
 \end{figure}

\section{Results}

\subsection{Hydrogen: 2-photon Emission and Broad Ly$\alpha$}

\subsubsection{\ion{H}{1} Continuum}  

Figure 3 shows that the emission and absorption lines are superposed upon a far-UV continuum with $F_{\lambda}$~$>$~10$^{-4}$ photons cm$^{-2}$ s$^{-1}$ \AA$^{-1}$.  This continuum was marginally detected in previous STIS G140L spectra, with the emission being concentrated in the ring plane. It is detected in our COS observations at high significance, ($Flux$ / $Error$) $\gtrsim$ 8 per spectral resolution element from 1360~--~1540~\AA.  

We can rule out the possibility that this emission is uncorrected scattered light from Star 3.  As described in Section 2, care was taken to isolate the emission from Star 3 at the bottom of the MCP detector (Figure 2).  At the observed continuum levels, the scattered light from Star 3 would have to be over twice as bright outside of the stellar extraction region as within it, and relatively constant with angular separation from the stellar spectrum.  Additionally, the observed continuum of Star 3 has a different spectral shape than the supernova continuum and there is no evidence for stellar features in the supernova spectrum (such as photospheric and wind absorption lines of \ion{C}{3} $\lambda$1175 and \ion{Si}{4} $\lambda$1394, 1403; Pellerin et al. 2002).\nocite{pellerin02} Therefore, we consider it unlikely that Star 3 contributes more than 10\% to the supernova continuum at $\lambda$~$>$~1350~\AA.  


We propose that this continuum is predominantly H I 2-photon (2$s$~$^{2}S_{1/2}$~--~1$s$~$^{2}S_{1/2}$)  emission.  We fit the 1350~$\leq$~$\lambda$~$\leq$~1720~\AA\ spectrum with the analytic formula of~\citet{nussbaumer84},
\begin{equation}
J_{\lambda}~=~\frac{1}{4\pi}\frac{hc}{\lambda^{3}} A(\lambda) N_{2s ^{2}S}  
\end{equation}
where $J_{\lambda}$ is the emissivity and $N_{2s ^{2}S}$ is the \ion{H}{1} column density in the $2s ^{2}S$ state.  This function rises sharply from 1216~\AA\ to a maximum at $\lambda$~$\sim$~1420~\AA, and then declines slowly to the red.   
The 2-photon spectral model is shown in orange in Figure 6.  Based on our near-UV images, we estimate that the circumstellar ring fills approximately 30~$\pm$~20\% of the inner 2\arcsec\ of the COS aperture where the transmission is $>$ 0.6.  Assuming this filling fraction, the continuum level determines the column density, and we find log$_{10}$($N_{2s ^{2}S}$(H)) = 21.34$^{+0.50}_{-0.26}$~cm$^{-2}$.  

The total integrated (1216~\AA~--~6 $\mu$m) \ion{H}{1} 2-photon flux is 9.1~$\pm$~0.6 $\times$~10$^{-1}$ photons cm$^{-2}$ s$^{-1}$, and the observed ratio of Ly$\alpha$ line emission to 2-photon is $F$(Ly$\alpha$)/$F$(2$s$) = 1.96~$\pm$~0.23.  Since this ratio is consistent with the 2$p$/2$s$ ratio of 2.1 (observed as $F$(Ly$\alpha$)/$F$(2$s$)) expected for recombination at $T$~$\approx$~10$^{4}$ K~\citep{spitzer78} and the (1$s$~$\rightarrow$~2$p$)/(1$s$~$\rightarrow$~2$s$) ratio of 2.05~--~2.10 expected for excitation by thermal electrons~\citep{callaway88}, we infer that it is likely that the Ly$\alpha$ and the 2-photon emission come from the same source.  

An important constraint on the origin of the far-UV continuum is the density of the emitting region.  \ion{H}{1}  2-photon emission can be suppressed by a factor $[1 + n_p/n_{cr}]^{-1}$ due to collisional depopulation of the 2$s$ state, where the critical density is given by $n_{cr} = A_{2\gamma}/C_{sp}$ and $n_p$ is the density of protons.  The (2$s$~$\rightarrow$~2$p$)  collision rate coefficient is $C_{sp}$~=~5.31~$\times$~10$^{-4}$ cm$^{3}$ s$^{-1}$~(Seaton 1955; see also Dennison et al. 2005 for a discussion of 2$s$ and 2$p$ level populations in \ion{H}{2} regions) and $A_{2\gamma}$~$\sim$~8.23 s$^{-1}$ for the 2-photon decay~\citep{klarsfeld69}, therefore $n_{cr}$ = 1.5~$\times$~10$^{4}$ cm$^{-3}$.\nocite{seaton55,dennison05}    

Based on the observed 2-photon flux, we can predict the total amount of associated H$\alpha$ if this emission is predominantly recombination. In Case B recombination, the rate of population of the $2s$ state is 3.3 times the rate of emission of H$\alpha$~\citep{osterbrock06}.  The expected H$\alpha$ in this scenario is $\approx$~2.8 $\times$~10$^{-1}$ photons cm$^{-2}$ s$^{-1}$.  We have made an estimate of the total H$\alpha$ flux from the circumstellar ring by analyzing STIS G750M spectra obtained in 2009 October ($HST$ exposure IDs OB7I200A0~--~OB7I200D0), making a correction for differential reddening between H$\alpha$ and the 2-photon emission. The total H$\alpha$ emission from the ring is $\approx$~2.5 $\times$~10$^{-1}$ photons cm$^{-2}$ s$^{-1}$, very close to the value expected if the 2-photon emission was produced by recombination.  However, \citet{groningsson08b} find that the H$\alpha$ flux from the ring is dominated by high-density ($>$~4~$\times$~10$^{6}$ cm$^{-3}$) shocked material around 5700 days after the explosion (October 2002). 2-photon emission could be highly suppressed from such gas, which has $n_p/n_{cr} \gtrsim$~ 300. 2-photon emission from lower density ($\lesssim$~5~$\times$~10$^{3}$ cm$^{-3}$; Matilla et al. 2010) gas in the unshocked ring would not be suppressed, but such gas can only account for $\sim$~ 20\% of the observed 2-photon emission.  

Because we observe no comparable source of H$\alpha$ emission, we conclude that the majority of the 2-photon emission is produced not by recombination, but by thermal ($T$~$\approx$~10$^{4}$ K) 
electron impact excitation of the $2s ^{2}S$ state of neutral hydrogen atoms in the outer ejecta, near the reverse shock front.  This low-density gas ($n_{H}$~$\sim$~100 cm$^{3}$; Smith et al. 2005; Heng et al. 2006) is heated by X-rays emitted from the shocked gas near the hotspots.  In particular, the energy deposition of the soft X-ray/EUV photons from the shocked hotspots will be concentrated in a layer near the reverse shock~(Fransson et al. 2011).
When the ionization fraction is $\gtrsim$~3~$\times$~10$^{-2}$, the majority of the X-ray energy  heats the gas through Coulomb stopping of fast photoelectrons~\citep{xu92,kozma92}.  The primary coolants for this gas are Ly$\alpha$ and 2-photon emission.

Is it reasonable to expect this level of X-ray heating of gas near the reverse shock?  We calculate the total X-ray flux from the circumstellar ring using the two-component model spectrum of~\citet{zhekov06}, scaled to the total 0.5~--~2.0 keV luminosity observed by $Chandra$ near day 8000 ($L$(0.5~--~2.0 keV)~$\approx$~1.5~$\times$~10$^{36}$ erg s$^{-1}$; Racusin et al. 2009).\nocite{racusin09}  The total 2-photon luminosity is $L$(2$s$)~$\approx$~2.2~$\times$~10$^{36}$ erg s$^{-1}$, therefore the 0.5~--~2.0 keV X-ray flux is insufficient to power the far-UV continuum.  However, the total shock luminosity is most likely dominated by emission in the soft X-ray/EUV band (0.01~--~0.5 keV) that is attenuated by the neutral hydrogen in the interstellar media of the Milky Way and LMC (Fransson et al. 2011).  The  luminosity in this band inferred from the model by Zhekov et al. is $L$(0.01~--~0.5 keV)~$\approx$~3~$\times$~10$^{38}$ erg s$^{-1}$.  Assuming that roughly half of this emission intersects the outer ejecta, we find that approximately~1.4\% of the soft X-ray/EUV luminosity from the shocked ring must be reprocessed into \ion{H}{1} 2-photon emission.   

\subsubsection{\ion{H}{1} Line Emission}  

Figures 4 and 5 show the broad emission from \ion{H}{1} Ly$\alpha$.  This emission was first observed in $HST$-STIS spectra 10.25 years after the SN~1987A explosion~\citep{sonneborn98}.   Subsequent studies have explored the reverse shock Ly$\alpha$ and H$\alpha$ emission in more detail~\citep{michael98,michael03}, including its brightening and deceleration (Heng et al. 2006; see also Smith et al. 2005 and Fransson et al. 2011 for discussion of the H$\alpha$ evolution).\nocite{smith05,heng06}  The combination of increased sensitivity and spectral resolution and low instrumental background of COS enable us to produce the highest-quality ultraviolet velocity profiles of the reverse shock emission to date.  The velocity distribution of the neutral hydrogen emission extends from $-$13000~--~+8000 km s$^{-1}$
(Figure 5).   The velocity maxima are much smaller than the initial observations by~Sonneborn et al. (1998; $\pm$~20000 km s$^{-1}$) and are consistent with the decrease in maximum projected velocity observed from 2004 to 2010~\citep{france10c}.  Using the limited angular resolution of COS, we confirm that emission at negative velocities is concentrated towards the northern side of the circumstellar ring while positive velocities are concentrated on the southern side of the ring.  
The total integrated ($-$12000~--~+8000 km s$^{-1}$) Ly$\alpha$ flux inferred from Figure 5
is 1.78~$\pm$~0.18 photons cm$^{-2}$ s$^{-1}$.   

\begin{figure}
 \begin{center}
 \epsfig{figure=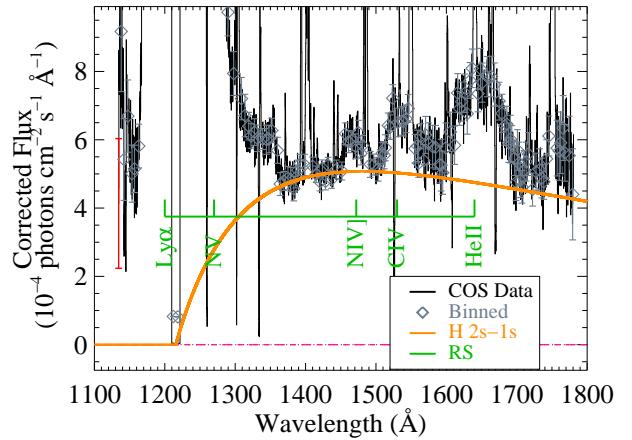,width=2.6in,angle=90}
 \caption{\label{cosdata} The combined WNW and ENE spectrum of SN~1987A ($black$) and the spectrum averaged over 0.8~\AA\ bins chosen to avoid narrow features from hotspot emission and interstellar absorption.   These data have been corrected for interstellar neutral hydrogen absorption (log$_{10}$(N(H))~=~21.43) and interstellar dust extinction, assuming $E(B-V)$~=~0.19 and $R_{V}$~=~3.1.  The broad spectral features are produced by a combination of hydrogen 2-photon emission (2$s$~$\rightarrow$~1$s$) and ionic emission from the reverse shock.  A theoretical hydrogen 2-photon spectrum is shown overplotted in orange.  
Reverse shock emission features are identified in green.  
  }
 \end{center}
 \end{figure}

Previous work noted that the Ly$\alpha$/H$\alpha$ ratios from the reverse shock exceed the 5:1 photon production ratio expected for a Balmer-dominated shock~\citep{heng06,heng07}.  \citet{france10c} found Ly$\alpha$/H$\alpha$ ratios $\geq$~30 from $-$8000~--~$-$2500 km s$^{-1}$ and $\geq$~20 from +3000~--~+7000 km s$^{-1}$ for isolated cuts across the northern and southern sides of the reverse shock, respectively.  They attributed the enhanced Ly$\alpha$ emission to a second source of Ly$\alpha$ photons.   They argued that Ly$\alpha$ photons from the hotspots are resonantly scattered by onrushing hydrogen with a distribution of velocities spanning a width $\Delta$$v_{H}$~$\sim$~3000~--~9000 km s$^{-1}$ (unlike Ly$\alpha$, H$\alpha$ is not a resonance line and therefore hotspot H$\alpha$ photons pass through the debris freely).  


While the Ly$\alpha$ enhancement at the largest negative velocities can likely be explained by this mechanism, our preceding discussion of the 2-photon continuum suggests that the majority of the lower velocity ($\pm$~7000 km s$^{-1}$) Ly$\alpha$ excess is attributable to the X-ray heating of the outer supernova debris.  As described above, Ly$\alpha$ is the primary coolant for mostly neutral hydrogen-rich gas excited by soft X-ray/EUV photoelectrons, and the agreement between the observed and theoretical $F$(Ly$\alpha$)/$F$(2$s$) ratio therefore argues that thermal electron collisions dominate the production of Ly$\alpha$ photons in the outer ejecta near the reverse shock front.  Resonant scattering will cause the newly produced Ly$\alpha$ photons to emerge preferentially in the outward direction, substantially favoring the blue-shifted wing as observed. 

\begin{deluxetable*}{lcccc}
\tabletypesize{\footnotesize}
\tablecaption{Continuum Subtracted SN~1987A Reverse Shock Emission \label{sn2010_cosdat} }
\tablehead{
\colhead{Species} & \colhead{$\Delta\lambda_{obs}$} & \colhead{$\Delta$$v$\tablenotemark{a}} & \colhead{Line Flux}    & \colhead{Flux Uncertainty} \\ 
    & (\AA) &  (km s$^{-1}$) & (photons cm$^{-2}$ s$^{-1}$)  &   (photons cm$^{-2}$ s$^{-1}$)
}
\startdata
\ion{H}{1}\tablenotemark{b}  & 1169.6~--~1208.0  &  -12000~--~-1500  &   1.1~$\times$~10$^{0}$  &   5.9~$\times$~10$^{-3}$ \\ 
\ion{H}{1}  & 1222.8~--~1247.1  &  1500~--~8000  &   4.5~$\times$~10$^{-1}$  &   4.1~$\times$~10$^{-3}$ \\ 

\tableline
\ion{N}{5}\tablenotemark{c}  & 1197.9~--~1233.9  &  -10000~--~-1000  &   5.7~$\times$~10$^{-2}$  &   6.2~$\times$~10$^{-3}$ \\ 
\ion{N}{5}  & 1247.1~--~1285.5  &  2000~--~12000  &   8.0~$\times$~10$^{-2}$  &   8.7~$\times$~10$^{-4}$ \\ 

\tableline
\ion{N}{4}]  & 1428.3~--~1478.8  &  -12000~--~-1500  &   1.8~$\times$~10$^{-3}$  &   4.9~$\times$~10$^{-4}$ \\ 
\ion{N}{4}]\tablenotemark{d}  & 1497.0~--~1523.8  &  1500~--~8000  &  $\cdots$  &   $\cdots$ \\ 

\tableline
\ion{C}{4}  & 1491.9~--~1539.4  &  -12000~--~-1500  &   4.6~$\times$~10$^{-3}$  &   5.4~$\times$~10$^{-4}$ \\ 
\ion{C}{4}  & 1558.9~--~1584.6  &  1500~--~8000  &   2.1~$\times$~10$^{-3}$  &   4.8~$\times$~10$^{-4}$ \\ 

\tableline
\ion{He}{2}  & 1590.4~--~1633.7  &  -10000~--~-1000  &   6.8~$\times$~10$^{-3}$  &   7.8~$\times$~10$^{-4}$ \\ 
\ion{He}{2}  & 1658.0~--~1705.0  &  2000~--~12000  &   9.6~$\times$~10$^{-3}$  &   8.1~$\times$~10$^{-4}$ 
\enddata
\tablenotetext{a}{$v$~=~0 assumed to be the rest wavelength of the stronger lines of the \ion{N}{5} and \ion{C}{4} doublets.} 
\tablenotetext{b}{Corrected for interstellar neutral hydrogen absorption of 
 log$_{10}$(N(H))~=~21.43, however the inner $\pm$~6~\AA\ ($\pm$~1500 km s$^{-1}$) 
 cannot be reconstructed. All reverse shock fluxes have been corrected for interstellar dust extinction, assuming $E(B-V)$~=~0.19 and $R_{V}$~=~3.1.}
\tablenotetext{c}{Blue \ion{N}{5} flux is overwhelmed by reverse shock Ly$\alpha$ emission and interstellar \ion{H}{1} absorption.  We assume the \ion{N}{5} Blue/Red ratio is the same as the \ion{He}{2} reverse shock emission (~$\approx$~0.7).
} 
\tablenotetext{d}{Red \ion{N}{4}] flux is obscured by the stronger blue emission from \ion{C}{4}.}
\end{deluxetable*}

\subsection{\ion{He}{2} $\lambda$1640 Emission}

The broad emission feature that we attribute to \ion{He}{2} $\lambda$1640 from the reverse shock is easily visible above the 1600~--~1700~\AA\ continuum in Figure 6.  In Figure 7, 
we compare the \ion{He}{2} velocity profile with the red side of the \ion{N}{5} $\lambda$1240 (\S4.3.1) profile.
As neutral or singly ionized helium crosses the reverse shock, collisions with thermal electrons~\citep{laming96} cause the helium to emit the analog to H$\alpha$.  
The velocity distribution of the ionized helium is expected to be quite different from that of neutral hydrogen because the He$^{+}$ ions that have crossed the reverse shock will be deflected by turbulent magnetic fields in the shock isotropization zone~\citep{michael98}.  The broad \ion{He}{2} displays a somewhat asymmetric line shape, spanning approximately $-$9000~--~+11000 km s$^{-1}$.   


Interpolating the integrated red and blue reverse shock \ion{He}{2}~$\lambda$1640 photon fluxes given in Table 2 through the narrow emission lines\footnote{The total emission from the reverse shock is interpolated through the central region lost to narrow-line emission.  In the case of \ion{He}{2}, $F^{RS}_{HeII}$~=~(1 + (3000/22000)) $\times$ ($F^{RS,red}_{HeII}$ + $F^{RS,blue}_{HeII}$).}, we measure a total flux of 1.9~$\times$~10$^{-2}$ photons cm$^{-2}$ s$^{-1}$.  
 Because this line is the hydrogenic analog to H$\alpha$, the $F$(1640)/$F$(H$\alpha$) ratio should give a direct measure of the relative abundance of helium (by number), assuming full hydrogen and helium ionization and that our line identification and continuum subtraction are correct.   We compare our \ion{He}{2} measurement with the H$\alpha$ flux from the day $\sim$~8000 VLT-UVES observations presented by Fransson et al. 2011, $F_{obs}$(H$\alpha$)~$\approx$~1.4 $\times$~10$^{-13}$ erg cm$^{-2}$ s$^{-1}$.  We convert this value into a total reverse shock H$\alpha$ by applying a factor of 1.86 slit correction, a factor of 1.45 reddening correction at H$\alpha$~\citep{groningsson08}, and a factor of 24/22 to account for the relative time of observation after the explosion.  The integrated H$\alpha$ photon flux in 2011 is~$\approx$~0.14 photons cm$^{-2}$ s$^{-1}$.  The He/H abundance ratio in the reverse shock is 0.14~$\pm$~0.06, in agreement with the He/H abundance ratio derived by~\citet{mattila10} for the circumstellar ring.  
The fact that the \ion{He}{2}~$\lambda$1640/H$\alpha$ ratio reproduces the circumstellar ring He/H abundance ratio suggests that the neutral hydrogen pre-ionization predicted by~\citet{smith05} is negligible 22 years after the explosion.  

\citet{borkowski97} predicted the time evolution of the observable (accounting for attenuation due to interstellar reddening) reverse shock flux in several abundant ions, including \ion{He}{2}.  Their predictions only extend to 2007, but we can extrapolate their curves to 24 years after the explosion for comparison with the COS data.  We estimate their prediction for the observed \ion{He}{2} flux in 2011 to be $\sim$ 1~--~3~$\times$~10$^{-3}$ photons cm$^{-2}$ s$^{-1}$ from Figure 4 of~\citet{borkowski97}.   Correcting these values by a factor of 4 to account for interstellar reddening, their prediction for the \ion{He}{2} flux is 0.4~--~1.2 $\times$~10$^{-2}$ photons cm$^{-2}$ s$^{-1}$, only a factor of $\sim$~2 below the \ion{He}{2} flux observed by COS.  Differences between the predicted and observed \ion{He}{2} fluxes are likely due to assumed electron temperatures that are lower than suggested by the observations (see the next subsection).

 \begin{figure}
 \begin{center}
 \epsfig{figure=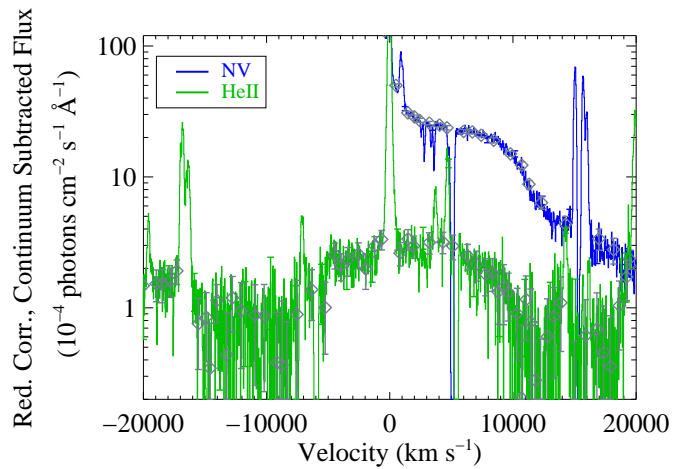,width=2.6in,angle=90}
 \caption{\label{cosdata} 2-photon continuum-subtracted \ion{N}{5} and \ion{He}{2} in the combined WNW + ESE spectra.
The blue side of the \ion{N}{5} distribution is lost under the RS Ly$\alpha$ emission.  The \ion{He}{2} velocity profile appears qualitatively similar to that of \ion{N}{5}. 
  }
 \end{center}
 \end{figure}

\subsection{Carbon and Nitrogen Emission Lines}

\subsubsection{\ion{N}{5} $\lambda$1240}

Prior to the first STIS observations of SN~1987A, Borkowski et al. (1997) predicted that strong reverse shock emission from Li-like \ion{N}{5}~ $\lambda$1240 would be detectable.  This emission was not apparent, however, in the first deep far-UV STIS spectra presented by \citet{sonneborn98} and \citet{michael98}. In recent (January 2010) STIS observations, we tentatively detected broad, redshifted \ion{N}{5} emission (Figure 3 of France et al. 2010), but low S/N precluded a detailed analysis.  Now, in our COS observations, we unambiguously detect this emission.  While the blue wing of the line is lost under the bright Ly$\alpha$ emission, we observe the complete red wing of the \ion{N}{5} $\lambda$1240 velocity profile.  In Figure 7, we compare the line profile of \ion{N}{5} with that of \ion{He}{2}, observing that the red wings of the two profiles are qualitatively similar.  Both ions present a boxy line profile, with a flat top and fall off between +9000~--~+10000 km s$^{-1}$.  The \ion{N}{5} profile extends to at least +14000 km s$^{-1}$, where the data are contaminated by the geocoronal \ion{O}{1} triplet.    

At the relevant energy scales of nonradiative supernova shocks, $v_{s}$~$\gtrsim$~10$^{3}$ km s$^{-1}$, the excitation cross sections for Li-like species (e.g., C$^{3+}$, N$^{4+}$, O$^{5+}$) by collisions with protons and heavier ions are considerably larger than the cross-section for ionization to their He-like stage~\citep{laming96,borkowski97}.  Therefore, these species may emit many (several hundred) line photons (predominantly \ion{C}{4} $\lambda$ 1550, \ion{N}{5} $\lambda$ 1240, \ion{O}{6} $\lambda$ 1032 for the ions listed above)  for every atom that crosses the reverse shock front before becoming ionized.  The total ratio of \ion{N}{5} $\lambda$1240 photons emitted to H$\alpha$ is given by 
\begin{equation}
\frac{F(NV)}{F(H\alpha)}~=~ \frac{1}{0.2}~\frac{x_{N}}{x_{H}}~\frac{R(1240)}{R(N^{4+} \rightarrow N^{5+})}
\end{equation}
where 0.2 is the number of H$\alpha$ photons emitted per neutral hydrogen atom crossing the shock~\citep{michael03}, $x_{N}/x_{H}$ is the nitrogen abundance ratio, $R(1240)$ is the rate of electron and ion impacts that result in \ion{N}{5} emission, and $R(N^{4+} \rightarrow N^{5+})$ is the rate of ionization to N$^{5+}$.  The relative nitrogen abundance in the circumstellar ring is 2.8~($\pm$~1.1)~$\times$~10$^{-4}$~\citep{mattila10}.  Given the similarity of the red side of the \ion{He}{2} and \ion{N}{5} profiles, we compute the total \ion{N}{5} reverse shock photon flux by assuming that the relative red/blue contribution is the same as \ion{He}{2} (Table 2) and interpolating through the narrow emission lines, as described above.  Under these assumptions, the observed $F$(\ion{N}{5})/$F$(H$\alpha$) ratio is $\approx$~1.15.   Folding in the nitrogen abundance, we can therefore make a rough estimate of the ratio $R(1240)$/$R(N^{4+} \rightarrow N^{5+})$ $\sim$~850$^{+550}_{-250}$ required to explain the observed line ratio.

$R(1240)$ can be evaluated analytically,
\begin{equation}
R(1240)~=~n_e~\langle \sigma_e v_{s} \rangle~+~n_H~\langle \sigma_p v_{s} \rangle~+~n_{He}~\langle \sigma_\alpha v_{s} \rangle
\end{equation}
where the cross-sections, $\sigma_{e,p,\alpha}$, are for excitation of \ion{N}{5} $\lambda$1240 by electrons, protons, and alpha particles, respectively. We simplify the averages over velocity distribution by evaluating the expression at a single shock velocity, $v_{s}$ = 10$^{4}$ km s$^{-1}$.  Cross-sections for the proton and alpha particle collisions are taken from Table 2 of~\citet{laming96} for proton energies ($E_{p}$~=~1/2~$m_{p}$$v_{s}^{2}$) of 522 keV (we assume the 544 keV values) and alpha particle energies ($E_{\alpha}$~=~1/2~(4$m_{p})$$v_{s}^{2}$) of 2088 keV (we assume 1.2 times the 2720 keV values).  The electron impact excitation cross-sections are not as well determined, but we expect them to be similar to those of the protons at high shock velocity.  Therefore, we estimate $\sigma_{e,p,\alpha}$ = [4.31, 4.31, 14.4]~$\times$~10$^{-17}$ cm$^{-2}$, respectively.   
Taking $n_{e}$ = 450 cm$^{-3}$~\citep{borkowski97}, $n_{H}$ $\approx$ 100 cm$^{-3}$~\citep{smith05,heng06}, and $n_{He}$ = 0.17 $n_{H}$~\citep{mattila10}, we compute a total \ion{N}{5} excitation rate for $v_{s}$~=~10$^{4}$ km s$^{-1}$; 
$R(1240)$ = 2.6~$\times$~10$^{-5}$ s$^{-1}$.  

Ionization rates by baryons are expected to be small compared to those by electrons due to the high energy of post-shock protons and heavier species~\citep{laming96}, therefore we neglect their contribution to the N$^{4+}$ ionization rates.  Cross-sections for the direct ionization to N$^{5+}$ by thermal electrons can be calculated as a function of electron energy from Equation 1 of~\citet{arnaud85}.  Using coefficients for the Li-sequence (their Table 1), we calculate an electron ionization cross-section of $\sigma^{NV}_{ion}$ = 1.5~$\times$~10$^{-18}$ cm$^{-2}$ at $E_{e}$~=~107 eV\footnote{$E_{e}$~=~$\frac{3}{16}$ $m_{e}$$v^{2}_{s}$}. 
The direct ionization rate, $R(N^{4+} \rightarrow N^{5+})$ = $n_{e}$~$\langle$ $\sigma^{NV}_{ion}$ $v_{s}$ $\rangle$ = 1.5~$\times$~10$^{-7}$ s$^{-1}$.
 This gives $R(1240)$/$R(N^{4+} \rightarrow N^{5+})$ $\approx$~170, which is outside the range allowed by the observations. 

If, on the other hand, there is partial equilibration between the electron and proton distributions~\citep{cargill88,laming96,heng10}, then $T^{'}_{e}$~=~$\beta_{eq}$$T_{p}$, where 
\begin{equation}
T_{p}~=~\frac{3}{16}~\frac{\mu m_{p} v_{s}^{2}}{k_B} 
\end{equation}
and $\beta_{eq}$ is the equilibration factor, which has a maximum value of 1.  $\mu$ is the mean particle weight, $\mu$~$\approx$~0.55 (for $x_{H}$ = 0.85, $x_{He}$ = 0.14, and $x_{Z}$ = 0.01).  
In this case $E^{'}_{e}$ = 1/2 $m_{e}$ (2$k_{B}$$T^{'}_{e}$ / $m_{e}$) = $k_{B}$$T^{'}_{e}$.  Using this energy to calculate the ionization cross-sections, we find that for  $\beta_{eq}$ = 0.14~--~0.35 ($T^{'}_{e}$~=~1.7~--~4.4~$\times$~10$^{8}$ K), the electron ionization cross-sections $\sigma^{NV}_{ion}$ = 9.7~--~4.2 $\times$~10$^{-20}$ cm$^{2}$, and the ionization rates of $R(N^{4+} \rightarrow N^{5+})$ 
= 4.3~--~1.9~$\times$~10$^{-8}$ s$^{-1}$.  This gives $R(1240)$/$R(N^{4+} \rightarrow N^{5+})$ $\sim$~610~--~1380, approximately the range required to account for the observed $F$(\ion{N}{5})/$F$(H$\alpha$) ratio\footnote{We note that if we assume that the ionization cross-sections scale as $\sigma_{ion}^{NV}$~$\propto$~$T_{e}^{'-0.5}$ in the high-$T_{e}$ limit, then the ion-electron equilibration factor could be a factor of 2~--~3 higher, approaching the limit of complete equilibration.}.  $v_{s}$~=~10$^{4}$ km s$^{-1}$ is used as the fiducial velocity in our calculations, but the general conclusions hold for a range of possible shock velocities.  Computing $R(1240)$/$R(N^{4+} \rightarrow N^{5+})$
for velocities (5~--~12)~$\times$~10$^{3}$ km s$^{-1}$, we find that $\beta_{eq}$ decreases by ~$\sim$~50\% across this range, but $\beta_{eq}$~$>$~0.1 is required to explain the data for all velocities considered.  

It seems clear that partial ion-electron equilibration is required to explain the observation of strong reverse shock \ion{N}{5} emission from SN~1987A. However, values of $\beta_{eq}$ = 0.14~--~0.35 are greater than those favored by~\citet{laming96} from their non-radiative shock modeling of SN1006. Moreover, equilibration values of $\beta_{eq}$~$\geq$~0.1 for the high shock velocity ($v_{s}$~$\sim$~10$^{4}$ km s$^{-1}$) in SN~1987A are inconsistent with some recent results on electron-ion equilibration in collisionless shocks. Ghavamian et al. (2007) discuss a sample of older ($\sim$~10$^{3}$ yr) supernova remnants whose electron-to-proton temperature ratio is inversely proportional to the square of the shock velocity, $\beta_{eq}$~$\propto$~$v_{s}^{-2}$; however, this relation has not been confirmed in subsequent work~\citep{helder11}.  Additionally, \citet{adelsberg08} present evidence suggesting an increasing $\beta_{eq}$ for  $v_{s}$~$>$~2000 km s$^{-1}$.  
Finally, we note that very high electron temperatures ($T_{e}$~$\sim$~10$^{9}$ K) were inferred for the high velocity ejecta in SN~1993J~\citep{fransson96b}, suggesting that electron heating may be efficient in high velocity environments (see \S5.3). 

We also compare the observed \ion{N}{5} flux with that predicted by~\citet{borkowski97}.  Extrapolating their prediction to the time of our observations, we would have expected (1.5~--~2.0)~$\times$~10$^{-13}$ erg cm$^{-2}$ s$^{-1}$ in 2011, or 0.06~--~0.08 photons cm$^{-2}$ s$^{-1}$ after including a factor of 6.3 to account for interstellar reddening.  We see that, similar to the prediction for \ion{He}{2}, these values are a factor of $\sim$~2 below their observed 2011 values ($F$(\ion{N}{5}) = 0.16 photons cm$^{-2}$ s$^{-1}$).  Overall, we consider the prediction made 14 years prior to the observation to be remarkably good.  The slight differences between the predicted and observed \ion{N}{5} flux can most likely be attributed to their choice of a value for $\beta_{eq}$ that is lower than we determine from the data.  Lower electron energies increase the ionization cross-section and hence $R(N^{4+} \rightarrow N^{5+})$, thereby reducing the total number of \ion{N}{5} photons emitted.

\subsubsection{\ion{N}{4}] $\lambda$1486}

Figure 6 shows that there is considerable substructure on the continuum in addition to the strong reverse shock emission from \ion{H}{1}, \ion{He}{2}, and \ion{N}{5}.  The weakest of these features spans $\sim$ 1430~--~1500~\AA.  We attribute this emission  to \ion{N}{4}] $\lambda$1486 from the reverse shock.  The \ion{N}{4}] emission is highly asymmetric about the rest velocity.  The blue-shifted component extends to roughly $-$8000 km s$^{-1}$, while the red side of the profile is at the noise level by +3000 km s$^{-1}$.  Additionally, the blue wing of the reverse shock \ion{C}{4} profile, discussed in the next section, overwhelms any additional red flux.   Again making the assumption that the red/blue \ion{N}{5} ratio is the same as that for \ion{He}{2}, and assuming that the red side of the \ion{N}{4}] profile contributes an additional 20\% to the total flux, we find that the total $F$(\ion{N}{5})/$F$(\ion{N}{4}]) ratio is ~$\sim$~72~$\pm$~17.  

\subsubsection{\ion{C}{4} $\lambda$1550}

Figure 6 also shows \ion{C}{4} $\lambda$1550 from the reverse shock covering $\sim$~1500~--~1580~\AA, with emission from $-$12000~--~$-$1500 km s$^{-1}$ having 2.2 times the total integrated photon flux as the 
+1500~--~+8000 km s$^{-1}$ interval (Table 2).  The red side of the velocity profile is only marginally above the noise level at $v_{CIV}$ $>$ +3000 km s$^{-1}$.  We consider the relative fluxes of \ion{C}{4} and \ion{N}{5} here, and in the next section we will discuss the relative velocity distributions of the reverse shock species.   

Interpolating the spectrum through the narrow emission lines, we measure an integrated broad line photon flux ($-$12000~--~+8000 km s$^{-1}$) of $F$(\ion{C}{4}) = 7.7 $\times$ 10$^{-3}$ photons cm$^{-2}$ s$^{-1}$.  With the previously-noted assumptions about the \ion{N}{5} flux in mind, we find a flux ratio of  $F$(\ion{N}{5})/$F$(\ion{C}{4})~$\approx$~20~$\pm$~3.   
The ratio of \ion{N}{5} $\lambda$1240/\ion{C}{4} $\lambda$1550 is given by an expression similar to equation (2):
\begin{equation}
\frac{F(NV)}{F(CIV)}~=~ \frac{x_{N}}{x_{C}}~\left[\frac{R(1240)}{R(N^{4+} \rightarrow N^{5+})}\right]~\left[\frac{R(1550)}{R(C^{3+} \rightarrow C^{4+})}\right]^{-1}
\end{equation}
The ratio of the quantities in square brackets is very close to 0.9 for $v_{s}$~=~(5~--~12)~$\times$~10$^{3}$ km s$^{-1}$ (the ratio is near unity because both are Li-like ions).   
 That implies that the abundance ratio of nitrogen/carbon atoms crossing the reverse shock is $x_{N}$/$x_{C}$~$\approx$~22~$\pm$~3, greater than the ratio $x_{N}$/$x_{C}$~$\approx$~8.5~$\pm$~3.5 inferred from spectroscopic observations of the equatorial ring (Lundqvist \& Fransson 1996; Mattila et al. 2010).

We note that while our estimate of the flux ratio \ion{N}{5}~$\lambda$1240/\ion{C}{4}~$\lambda$1550 is somewhat uncertain because our choice of the blue side of the \ion{N}{5} profile is speculative, the integrated flux in the red side of the \ion{N}{5} profile alone is more than 10 times greater than the $total$ \ion{C}{4} flux.  Therefore, we are confident that the \ion{N}{5}/\ion{C}{4} ratio is enhanced well beyond what can be attributed to the circumstellar ring abundances of the two species.  We return to this point in Section 5.4. 

 \begin{figure}
 \begin{center}
 \epsfig{figure=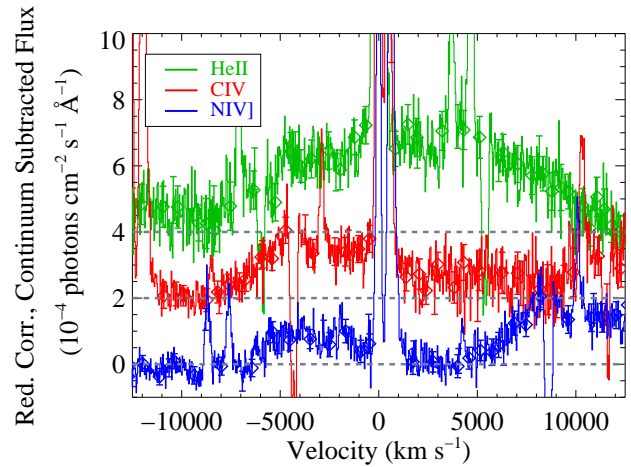,width=2.6in,angle=90}
 \caption{\label{cosdata} 2-photon continuum subtracted \ion{N}{4}], \ion{C}{4}, and \ion{He}{2} in the combined WNW + ESE spectra.    \ion{He}{2} and \ion{C}{4} have been offset by 4.0 and 2.0 photons cm$^{-2}$ s$^{-1}$ \AA$^{-1}$, respectively, for display purposes.  The zero-flux levels for all three ions are shown as the dashed gray line.  
  }
 \end{center}
 \end{figure}

\section{Discussion}

\subsection{Relative Velocity Distributions}

In Sections 4.2 and 4.3, we describe the velocity distributions of \ion{He}{2}, \ion{N}{5}, \ion{N}{4}], and \ion{C}{4}.  The red side of the \ion{N}{5} velocity profile is compared with \ion{He}{2} in Figure 7 and a comparison of the \ion{He}{2}, \ion{N}{4}], and \ion{C}{4} velocity profiles is shown in Figure 8.  
We observe qualitatively different velocity profiles from \ion{He}{2}, \ion{N}{4}], and \ion{C}{4}. \ion{He}{2} displays a mildly asymmetric profile spanning $\sim$~$-$9000~--~+11000 km s$^{-1}$, while \ion{N}{4}] and \ion{C}{4} are strongly weighted towards the blue, spanning $\sim$~$-$(9000~--~8000)~--~+3000 km s$^{-1}$.  Peculiar geometric projection of the velocity can be ruled out because all three ions were observed with the G160M instrument mode, with a constant spacecraft roll orientation.  

In nonradiative shocks, \ion{He}{2} is thought to be excited by collisions with electrons, while the Li-like species (\ion{N}{5} and \ion{C}{4}) are excited by ion-collisions~\citep{laming96}.  As we have shown above, the electron and ion distributions have likely not equilibrated in the SN~1987A debris, therefore it may be that the relative distributions of the impacting species is reflected in the velocity profiles of the excited ions.  One might also speculate that if ion collisions dominate the emission from metals, the preferentially blue \ion{N}{4}] and \ion{C}{4} profiles would reflect an asymmetric spatial distribution of the protons and ions.  The \ion{N}{4}] and \ion{C}{4} profiles also show spatial variations between the WNW and ENE spectra.  
In contrast, the \ion{He}{2} profile, which is primarily excited by electrons, is more spatially and spectrally symmetric.  This may indicate a more uniform distribution of electrons.  
With these possibilities in mind, 
one is tempted to separate the metals from the helium, however we observe \ion{N}{5} extending to red velocities as high as +10000 km s$^{-1}$ before declining.  Modeling of both the fluxes and velocity distributions of ions in the SN~1987A reverse shock would be very useful for our understanding of the observed velocity profiles. 

\subsection{Excess Emission at $\lambda$~$<$~1350~\AA}

The high throughput and low instrumental background of COS have allowed us to study the ultraviolet continuum of SN 1987A for the first time.  Similarly, we have presented the first conclusive observation of elements heavier than hydrogen in the reverse shock.  
Figures 3 and 6 show that there is additional ultraviolet continuum at $\lambda$~$<$~1350~\AA, above which we propose that hydrogen 2-photon emission can adequately explain the data.  As we discuss in \S4.1.2, the spectral contamination from Star 3 is inversely proportional to wavelength, as the cross-dispersion astigmatism height is largest at the shortest wavelengths.  This could account for some, but probably not all, of the bluest continuum emission.  Instead, we propose four speculative possibilities for this emission.  He$^{+}$ has a 2$p$~$\rightarrow$~1$s$ Ly$\alpha$ line at 304~\AA, and will emit a 2$s$~$\rightarrow$~1$s$ 2-photon continuum which peaks at $\lambda$~$\approx$~355~\AA, and declines to the red.  We estimate that this emission contributes $\approx$~10\% of the 1150~\AA\ continuum.   

There may be an additional Ly$\alpha$ component at very high velocities, although this would require $v_{Ly\alpha}$~$\gtrsim$~30~$\times$~10$^{3}$ km s$^{-1}$ ($\gtrsim$~0.1~$c$) to account for the reddest observed emission.  Another possibility is that low-ionization line emission from the core, as opposed to higher-ionization emission from atoms crossing the reverse shock, contributes to the $\lambda$~$\leq$~1350\AA\ emission.  \citet{jerkstrand11} describe 2-photon emission, scattered by low-ionization, high-opacity metal lines in the core, contributing to the observed spectrum at $\sim$~8 years following the explosion.  The observed excess begins at roughly the wavelength of the \ion{O}{1}]~$\lambda$1356~\AA\ emission line.  If an appreciable amount of low-ionization metal emission escapes from the near side of the core , it may be responsible for a portion of the observed flux.  

Perhaps the most likely scenario is one in which several minor reverse shock species contribute to the observed continuum between 1140~--~1350~\AA.  We have seen that reverse shock emission from Be-like ions (e.g. NIV] $\lambda$1486) are present.  Therefore, emission from lines of \ion{C}{3} $\lambda$1175 and \ion{O}{5}~$\lambda$1371 may also contribute\footnote{\ion{O}{5} is the next ionization stage down from the Li-like \ion{O}{6} $\lambda$1032 line which we do not detect, but is presumably in the reverse shock spectrum at shorter wavelengths.}. 

\subsection{Partial Ion-Electron Equilibration in SN~1987A: Evidence for the Cross-Shock Potential? }

In Section 4.3.1, we found that large ion-electron temperature equilibration ratios ($\beta_{eq}$ $\approx$~0.14~--~0.35) are required to explain the high \ion{N}{5} $\lambda$1240 flux observed in the spectrum of SN~1987A.~\citet{ghavamian07} predict equilibration ratios several times less than this for shock velocities observed in SN~1987A ($v_{s}$~$\sim$~10$^{4}$ km s$^{-1}$).  Therefore, we consider additional electron heating mechanisms that could explain the observed \ion{N}{5} emission.  One mechanism that has been proposed for collisionless shocks (usually assumed to be quasi-perpendicular) at low Alfv\'en mach numbers ($M_A$) is the cross-shock potential \citep[e.g][]{balikhin93}.  But 
the cross-shock potential is expected to become less important as $M_A$ increases because the plasma resistivity causes insufficient dissipation and the shock becomes unsteady~\citep{edmiston84}.  Electrons undergoing $E$ $\times$ $B$ drift across the shock  front are not guaranteed to see the same cross-shock potential. They may lose energy as well as gain it, and this limits the electron heating.  

However, it has also been suggested that at Alfven Mach numbers appropriate to shocks in supernova remnants, the shock transition becomes thin. The shock transition can develop a length scale less than the electron convective
gyroradius, thus eliminating any $E$ $\times$ $B$ drift within the shock. This onsets at $M_A$ = $\sqrt{m_{i}/m_{e}}$, and electron heating by the cross shock potential again becomes viable~\citep{balikhin93,gedalin08}. 
Typical SNR shocks have $M_A$ much higher than this, neglecting any modification of the preshock medium. 
Yet the predicted heating is not observed.  Most likely, $M_A$ at these shocks is much lower than expected due to amplification of the magnetic field by cosmic rays streaming ahead of the shock  ~\citep{bell04,bell05}. 
Electron heating in these systems occurs via plasma waves excited upstream by 
shock reflected ions or cosmic rays. At higher velocity shocks, or for those likely to have weak upstream magnetic field due to their environment and/or insignificant cosmic ray populations, $M_A$ $>$ $\sqrt{m_{i}/m_{e}}$ and the cross-shock
potential may be capable of heating electrons to the required energy. Therefore
the energetic environment of SN 1987A, especially at its reverse shock, may enable
additional electron heating mechanisms that do not contribute in lower $M_A$ systems.

\subsection{The \ion{N}{5}/\ion{C}{4} Ratio}

The final unresolved issue is the large \ion{N}{5}/\ion{C}{4} ratio in the reverse shock emission.  Carbon and nitrogen abundance ratios in the circumstellar ring suggest this number should be $\approx$~8~\citep{lundqvist96,mattila10}, however the observed flux ratios indicate N/C~$\approx$~22.  It may be that the N/C abundance ratio in the outer envelope of the progenitor was stratified prior to the ejection of the circumstellar rings, and that we are now seeing the first observational evidence of that stratification.  
A second possibility to account for this discrepancy is that ongoing thermonuclear processing continued to convert C to N in the supernova progenitor following the circumstellar ring ejection.  The CNO bi-cycle will enrich the $^{14}$N abundance at the expense of the abundances of $^{12}$C and $^{16}$O and, in equilibrium, will convert almost all of the primordial C and O into N~\citep{caughlan62}.   CNO processing has been invoked to explain the fact that the observed nitrogen abundance in the circumstellar ring is elevated by a factor $\sim$~10 over its value in the LMC~\citep{fransson89}.  Our observation that the He abundance does not change between the ring and reverse shock is qualitatively consistent with the 14E1 model presented by~\citet{shigeyama90}.  They show that the high-velocity material we observe crossing the reverse shock front is only a small fraction of the total ejected mass, and does not probe deep enough into the interior of the ejecta to observe significantly elevated He abundances.  

If CNO processing continued near the stellar surface following the ejection of the circumstellar rings, it could have in principle converted most of the remaining C and O abundances seen in the equatorial ring into N.  \citet{heng08} note reduced oxygen abundances, possibly related to the high \ion{N}{5}/\ion{C}{4} ratio observed in our observations.   For this explanation to be viable, the timescale to reach equilibrium in the CNO cycle must be $\lesssim$ 20,000 years, the interval since the ejection of the equatorial ring.  This condition will be met if the temperature of the shell where CNO burning takes place is $T$~$\gtrsim$~3.5~$\times$~10$^{7}$ K  (Caughlan and Fowler 1962, Table 4).
This effect could not only account for the high N/C ratio in the gas crossing the reverse shock, it could also increase the N/H ratio by an additional factor of $\sim$~2.  This would lower the required $\beta_{eq}$ derived in \S4.3.1.  


\section{Conclusions}

We have presented an analysis of deep $HST$-COS spectroscopy of SN~1987A.   Below, we summarize our primary results. 

\begin{enumerate}  
	\item We observe narrow lines from shocked gas in the circumstellar ring, broad emission lines from the reverse shock, and a strong detection of ultraviolet continuum emission.  Several of the emission line species and the far-UV continuum are conclusively detected for the first time. 
	\item The asymmetry in Ly$\alpha$ profile and its enhancement relative to H$\alpha$ suggests that most of the Ly$\alpha$ emission is a result of the illumination of the outer supernova debris by soft ($E$~$\leq$~0.5 keV) X-rays emitted by the shocked circumstellar ring.  
	\item The $\lambda$~$\gtrsim$~1350~\AA\ continuum may be described by hydrogen 2-photon emission originating in the outer ejecta, near the reverse shock front.  
	\item We present resolved velocity profiles of \ion{He}{2}, \ion{C}{4}, \ion{N}{4}], and \ion{N}{5} from the reverse shock.  A comparison of the velocity integrated \ion{He}{2} and H$\alpha$ velocity profile indicates a He/H abundance ratio of 0.14~$\pm$~0.06.  
In order to reproduce the observed $F$(\ion{N}{5})/$F$(H$\alpha$) line ratio, partial ion-electron equilibration is required, $T^{'}_{e}$~=~$\beta_{eq}$$T_{p}$, with $\beta_{eq}$ $\approx$~0.14~--~0.35.  Large values of $\beta_{eq}$ may be explained by additional electron heating by the cross-shock potential.
	\item The velocity profiles of \ion{C}{4} and \ion{N}{4}] are significantly different from that of \ion{He}{2}, which may be related to the different excitation processes (ion vs. electron collisions) for the different species.
	\item We observe additional continuum emission at $\lambda$~$<$~1350~\AA\ that is not readily explained.  We favor a scenario where several weaker emission lines contribute, but spectral overlap with the much stronger Ly$\alpha$ and \ion{N}{5} emission profiles prevents an unambiguous interpretation.  
	\item Finally, we observe that the \ion{C}{4} reverse shock emission is weaker than would be expected assuming circumstellar ring abundances.  This may be explained by chemical stratification in the outer envelope of the progenitor, and may indicate additional CNO processing between the period of circumstellar ring ejection and the supernova explosion.  
\end{enumerate}


 \acknowledgments
We thank Svetozar Zhekov for making his X-ray shock model available, and KF and SVP thank James Green for enjoyable discussions about the spectroscopic imaging capabilities of COS.  We thank Dave Arnett for helpful advice regarding progenitor envelope structure.  
This work was supported by NASA grants NNX08AC146 and NAS5-98043 to the 
University of Colorado at Boulder.  Data were obtained as part of $HST$ program GO 12241. Support for program GO-12241 was provided by NASA through a�grant from the Space Telescope Science Institute, which is operated by the�Association of Universities for Research in Astronomy, Inc., under NASA 
contract NAS5-26555.  

{\bf Author Affiliations:} \\
\footnotesize
1~--~Center for Astrophysics and Space Astronomy, University of Colorado, 389 UCB, Boulder, CO 80309; kevin.france@colorado.edu \\
2~--~JILA, University of Colorado and NIST, 440 UCB, Boulder, CO 80309 \\
3~--~Harvard-Smithsonian Center for Astrophysics, 60 Garden Street, MS-78, Cambridge, MA 02138, USA \\
4~--~Naval Research Laboratory, Code 7674L, Washington, DC 20375, USA \\
5~--~Service d'Astrophysique DSM/IRFU/SAp CEA - Saclay, Orme des Merisiers, FR 91191 Gif-sur-Yvette, France \\
6~--~Department of Astronomy, University of Virginia, P.O. Box 400325, Charlottesville, VA 22904-4325, USA \\
7~--~Department of Astronomy, The Oskar Klein Centre, Stockholm University, 106 91 Stockholm, Sweden \\
8~--~225 Nieuwland Science, University of Notre Dame, Notre Dame, IN 46556-5670, USA \\
9~--~ETH Z\"{u}rich, Institute for Astronomy, Wolfgang-Pauli-Strasse 27, CH-8093, Z\"{u}rich, Switzerland \\
10~--~Department of Physics and Astronomy, Hofstra University, Hempstead, NY 11549, USA \\
11~--~Space Telescope Science Institute, 3700 San Martin Drive, Baltimore, MD 21218, USA \\
12~--~INAF/CT, Osservatorio Astrofisico di Catania, Via S. Sofia 78, I-95123 Catania, Italy \\
13~--~Supernova Ltd, OYV \#131, Northsound Road, Virgin Gorda, British Virgin Islands \\
14~--~Department of Physics, University of Hong Kong, Pok Fu Lam Road, Hong Kong, China \\
15~--~Steward Observatory, University of Arizona, 933 North Cherry Avenue, Tucson, AZ 85721, USA \\
16~--~NASA Goddard Space Flight Center, Code 665, Greenbelt, MD 20771, USA \\
17~--~Department of Physics and Astronomy, Goucher College, 1021 Dulaney Valley Road, Baltimore, MD 21204, USA \\
18~--~Department of Astronomy, University of Texas, Austin, TX 78712-0259, USA \\
\normalsize


\bibliography{ms_emapj_cos87a}

\begin{thebibliography}{58}
\expandafter\ifx\csname natexlab\endcsname\relax\def\natexlab#1{#1}\fi

\bibitem[{{Arnaud} \& {Rothenflug}(1985)}]{arnaud85}
{Arnaud}, M. \& {Rothenflug}, R. 1985, \aaps, 60, 425

\bibitem[{{Balikhin} {et~al.}(1993){Balikhin}, {Gedalin}, \&
  {Petrukovich}}]{balikhin93}
{Balikhin}, M., {Gedalin}, M., \& {Petrukovich}, A. 1993, Physical Review
  Letters, 70, 1259

\bibitem[{{Bell}(2004)}]{bell04}
{Bell}, A.~R. 2004, \mnras, 353, 550

\bibitem[{{Bell}(2005)}]{bell05}
---. 2005, \mnras, 358, 181

\bibitem[{{Borkowski} {et~al.}(1997){Borkowski}, {Blondin}, \&
  {McCray}}]{borkowski97}
{Borkowski}, K.~J., {Blondin}, J.~M., \& {McCray}, R. 1997, \apjl, 476, L31+

\bibitem[{{Callaway}(1988)}]{callaway88}
{Callaway}, J. 1988, \pra, 37, 3692

\bibitem[{{Cardelli} {et~al.}(1989){Cardelli}, {Clayton}, \& {Mathis}}]{ccm}
{Cardelli}, J.~A., {Clayton}, G.~C., \& {Mathis}, J.~S. 1989, \apj, 345, 245

\bibitem[{{Cargill} \& {Papadopoulos}(1988)}]{cargill88}
{Cargill}, P.~J. \& {Papadopoulos}, K. 1988, \apjl, 329, L29

\bibitem[{{Caughlan} \& {Fowler}(1962)}]{caughlan62}
{Caughlan}, G.~R. \& {Fowler}, W.~A. 1962, \apj, 136, 453

\bibitem[{{Danforth} {et~al.}(2010){Danforth}, {Keeney}, {Stocke}, {Shull}, \&
  {Yao}}]{danforth10}
{Danforth}, C.~W., {Keeney}, B.~A., {Stocke}, J.~T., {Shull}, J.~M., \& {Yao},
  Y. 2010, \apj, 720, 976

\bibitem[{{Dennison} {et~al.}(2005){Dennison}, {Turner}, \&
  {Minter}}]{dennison05}
{Dennison}, B., {Turner}, B.~E., \& {Minter}, A.~H. 2005, \apj, 633, 309

\bibitem[{{Edmiston} \& {Kennel}(1984)}]{edmiston84}
{Edmiston}, J.~P. \& {Kennel}, C.~F. 1984, Journal of Plasma Physics, 32, 429

\bibitem[{{Fitzpatrick} \& {Walborn}(1990)}]{fitzpatrick90}
{Fitzpatrick}, E.~L. \& {Walborn}, N.~R. 1990, \aj, 99, 1483

\bibitem[{{France} {et~al.}(2009){France}, {Beasley}, {Keeney}, {Danforth},
  {Froning}, {Green}, \& {Shull}}]{france09}
{France}, K., {Beasley}, M., {Keeney}, B.~A., {Danforth}, C.~W., {Froning},
  C.~S., {Green}, J.~C., \& {Shull}, J.~M. 2009, \apjl, 707, L27

\bibitem[{{France} {et~al.}(2010){France}, {McCray}, {Heng}, {Kirshner},
  {Challis}, {Bouchet}, {Crotts}, {Dwek}, {Fransson}, {Garnavich}, {Larsson},
  {Lawrence}, {Lundqvist}, {Panagia}, {Pun}, {Smith}, {Sollerman}, {Sonneborn},
  {Stocke}, {Wang}, \& {Wheeler}}]{france10c}
{France}, K., {McCray}, R., {Heng}, K., {Kirshner}, R.~P., {Challis}, P.,
  {Bouchet}, P., {Crotts}, A., {Dwek}, E., {Fransson}, C., {Garnavich}, P.~M.,
  {Larsson}, J., {Lawrence}, S.~S., {Lundqvist}, P., {Panagia}, N., {Pun},
  C.~S.~J., {Smith}, N., {Sollerman}, J., {Sonneborn}, G., {Stocke}, J.~T.,
  {Wang}, L., \& {Wheeler}, J.~C. 2010, Science, 329, 1624

\bibitem[{{Fransson} \& al.(2011)}]{fransson11}
{Fransson}, C. \& al., e. 2011, \aap, 1

\bibitem[{{Fransson} {et~al.}(1989){Fransson}, {Cassatella}, {Gilmozzi},
  {Kirshner}, {Panagia}, {Sonneborn}, \& {Wamsteker}}]{fransson89}
{Fransson}, C., {Cassatella}, A., {Gilmozzi}, R., {Kirshner}, R.~P., {Panagia},
  N., {Sonneborn}, G., \& {Wamsteker}, W. 1989, \apj, 336, 429

\bibitem[{{Fransson} {et~al.}(1996){Fransson}, {Lundqvist}, \&
  {Chevalier}}]{fransson96b}
{Fransson}, C., {Lundqvist}, P., \& {Chevalier}, R.~A. 1996, \apj, 461, 993

\bibitem[{{Gedalin} {et~al.}(2008){Gedalin}, {Balikhin}, \&
  {Eichler}}]{gedalin08}
{Gedalin}, M., {Balikhin}, M.~A., \& {Eichler}, D. 2008, \pre, 77, 026403

\bibitem[{{Ghavamian} {et~al.}(2007){Ghavamian}, {Laming}, \&
  {Rakowski}}]{ghavamian07}
{Ghavamian}, P., {Laming}, J.~M., \& {Rakowski}, C.~E. 2007, \apjl, 654, L69

\bibitem[{{Gilmozzi} {et~al.}(1987){Gilmozzi}, {Cassatella}, {Clavel},
  {Fransson}, {Gonzalez}, {Gry}, {Panagia}, {Talavera}, \&
  {Wamsteker}}]{gilmozzi87}
{Gilmozzi}, R., {Cassatella}, A., {Clavel}, J., {Fransson}, C., {Gonzalez}, R.,
  {Gry}, C., {Panagia}, N., {Talavera}, A., \& {Wamsteker}, W. 1987, \nat, 328,
  318

\bibitem[{{Gordon} {et~al.}(2003){Gordon}, {Clayton}, {Misselt}, {Landolt}, \&
  {Wolff}}]{gordon03}
{Gordon}, K.~D., {Clayton}, G.~C., {Misselt}, K.~A., {Landolt}, A.~U., \&
  {Wolff}, M.~J. 2003, \apj, 594, 279

\bibitem[{{Gr{\"o}ningsson} {et~al.}(2008{\natexlab{a}}){Gr{\"o}ningsson},
  {Fransson}, {Leibundgut}, {Lundqvist}, {Challis}, {Chevalier}, \&
  {Spyromilio}}]{groningsson08}
{Gr{\"o}ningsson}, P., {Fransson}, C., {Leibundgut}, B., {Lundqvist}, P.,
  {Challis}, P., {Chevalier}, R.~A., \& {Spyromilio}, J. 2008{\natexlab{a}},
  \aap, 492, 481

\bibitem[{{Gr{\"o}ningsson} {et~al.}(2008{\natexlab{b}}){Gr{\"o}ningsson},
  {Fransson}, {Lundqvist}, {Lundqvist}, {Leibundgut}, {Spyromilio},
  {Chevalier}, {Gilmozzi}, {Kj{\ae}r}, {Mattila}, \&
  {Sollerman}}]{groningsson08b}
{Gr{\"o}ningsson}, P., {Fransson}, C., {Lundqvist}, P., {Lundqvist}, N.,
  {Leibundgut}, B., {Spyromilio}, J., {Chevalier}, R.~A., {Gilmozzi}, R.,
  {Kj{\ae}r}, K., {Mattila}, S., \& {Sollerman}, J. 2008{\natexlab{b}}, \aap,
  479, 761

\bibitem[{{Helder} {et~al.}(2011){Helder}, {Vink}, \& {Bassa}}]{helder11}
{Helder}, E.~A., {Vink}, J., \& {Bassa}, C.~G. 2011, \apj, 737, 85

\bibitem[{{Heng}(2010)}]{heng10}
{Heng}, K. 2010, PASA, 27, 23

\bibitem[{{Heng} {et~al.}(2008){Heng}, {Haberl}, {Aschenbach}, \&
  {Hasinger}}]{heng08}
{Heng}, K., {Haberl}, F., {Aschenbach}, B., \& {Hasinger}, G. 2008, \apj, 676,
  361

\bibitem[{{Heng} \& {McCray}(2007)}]{heng07}
{Heng}, K. \& {McCray}, R. 2007, \apj, 654, 923

\bibitem[{{Heng} {et~al.}(2006){Heng}, {McCray}, {Zhekov}, {Challis},
  {Chevalier}, {Crotts}, {Fransson}, {Garnavich}, {Kirshner}, {Lawrence},
  {Lundqvist}, {Panagia}, {Pun}, {Smith}, {Sollerman}, \& {Wang}}]{heng06}
{Heng}, K., {McCray}, R., {Zhekov}, S.~A., {Challis}, P.~M., {Chevalier},
  R.~A., {Crotts}, A.~P.~S., {Fransson}, C., {Garnavich}, P., {Kirshner},
  R.~P., {Lawrence}, S.~S., {Lundqvist}, P., {Panagia}, N., {Pun}, C.~S.~J.,
  {Smith}, N., {Sollerman}, J., \& {Wang}, L. 2006, \apj, 644, 959

\bibitem[{{Jerkstrand} {et~al.}(2011){Jerkstrand}, {Fransson}, \&
  {Kozma}}]{jerkstrand11}
{Jerkstrand}, A., {Fransson}, C., \& {Kozma}, C. 2011, \aap, 530, A45+

\bibitem[{{Klarsfeld}(1969)}]{klarsfeld69}
{Klarsfeld}, S. 1969, Physics Letters A, 30, 382

\bibitem[{{Kozma} \& {Fransson}(1992)}]{kozma92}
{Kozma}, C. \& {Fransson}, C. 1992, \apj, 390, 602

\bibitem[{{Laming} {et~al.}(1996){Laming}, {Raymond}, {McLaughlin}, \&
  {Blair}}]{laming96}
{Laming}, J.~M., {Raymond}, J.~C., {McLaughlin}, B.~M., \& {Blair}, W.~P. 1996,
  \apj, 472, 267

\bibitem[{{Larsson} {et~al.}(2011){Larsson}, {Fransson}, {{\"O}stlin},
  {Gr{\"o}ningsson}, {Jerkstrand}, {Kozma}, {Sollerman}, {Challis}, {Kirshner},
  {Chevalier}, {Heng}, {McCray}, {Suntzeff}, {Bouchet}, {Crotts}, {Danziger},
  {Dwek}, {France}, {Garnavich}, {Lawrence}, {Leibundgut}, {Lundqvist},
  {Panagia}, {Pun}, {Smith}, {Sonneborn}, {Wang}, \& {Wheeler}}]{larsson11}
{Larsson}, J., {Fransson}, C., {{\"O}stlin}, G., {Gr{\"o}ningsson}, P.,
  {Jerkstrand}, A., {Kozma}, C., {Sollerman}, J., {Challis}, P., {Kirshner},
  R.~P., {Chevalier}, R.~A., {Heng}, K., {McCray}, R., {Suntzeff}, N.~B.,
  {Bouchet}, P., {Crotts}, A., {Danziger}, J., {Dwek}, E., {France}, K.,
  {Garnavich}, P.~M., {Lawrence}, S.~S., {Leibundgut}, B., {Lundqvist}, P.,
  {Panagia}, N., {Pun}, C.~S.~J., {Smith}, N., {Sonneborn}, G., {Wang}, L., \&
  {Wheeler}, J.~C. 2011, \nat, 474, 484

\bibitem[{{Lawrence} {et~al.}(2000){Lawrence}, {Sugerman}, {Bouchet}, {Crotts},
  {Uglesich}, \& {Heathcote}}]{lawrence00}
{Lawrence}, S.~S., {Sugerman}, B.~E., {Bouchet}, P., {Crotts}, A.~P.~S.,
  {Uglesich}, R., \& {Heathcote}, S. 2000, \apjl, 537, L123

\bibitem[{{Lundqvist} \& {Fransson}(1996)}]{lundqvist96}
{Lundqvist}, P. \& {Fransson}, C. 1996, \apj, 464, 924

\bibitem[{{Matsuura} {et~al.}(2011){Matsuura}, {Dwek}, {Meixner}, {Otsuka},
  {Babler}, {Barlow}, {Roman-Duval}, {Engelbracht}, {Sandstrom},
  {Laki{\'c}evi{\'c}}, {van Loon}, {Sonneborn}, {Clayton}, {Long}, {Lundqvist},
  {Nozawa}, {Gordon}, {Hony}, {Panuzzo}, {Okumura}, {Misselt}, {Montiel}, \&
  {Sauvage}}]{matsuura11}
{Matsuura}, M., {Dwek}, E., {Meixner}, M., {Otsuka}, M., {Babler}, B.,
  {Barlow}, M.~J., {Roman-Duval}, J., {Engelbracht}, C., {Sandstrom}, K.,
  {Laki{\'c}evi{\'c}}, M., {van Loon}, J.~T., {Sonneborn}, G., {Clayton},
  G.~C., {Long}, K.~S., {Lundqvist}, P., {Nozawa}, T., {Gordon}, K.~D., {Hony},
  S., {Panuzzo}, P., {Okumura}, K., {Misselt}, K.~A., {Montiel}, E., \&
  {Sauvage}, M. 2011, Science, 333, 1258

\bibitem[{{Mattila} {et~al.}(2010){Mattila}, {Lundqvist}, {Gr{\"o}ningsson},
  {Meikle}, {Stathakis}, {Fransson}, \& {Cannon}}]{mattila10}
{Mattila}, S., {Lundqvist}, P., {Gr{\"o}ningsson}, P., {Meikle}, P.,
  {Stathakis}, R., {Fransson}, C., \& {Cannon}, R. 2010, \apj, 717, 1140

\bibitem[{{Michael} {et~al.}(1998){Michael}, {McCray}, {Borkowski}, {Pun}, \&
  {Sonneborn}}]{michael98}
{Michael}, E., {McCray}, R., {Borkowski}, K.~J., {Pun}, C.~S.~J., \&
  {Sonneborn}, G. 1998, \apjl, 492, L143+

\bibitem[{{Michael} {et~al.}(2003){Michael}, {McCray}, {Chevalier},
  {Filippenko}, {Lundqvist}, {Challis}, {Sugerman}, {Lawrence}, {Pun},
  {Garnavich}, {Kirshner}, {Crotts}, {Fransson}, {Li}, {Panagia}, {Phillips},
  {Schmidt}, {Sonneborn}, {Suntzeff}, {Wang}, \& {Wheeler}}]{michael03}
{Michael}, E., {McCray}, R., {Chevalier}, R., {Filippenko}, A.~V., {Lundqvist},
  P., {Challis}, P., {Sugerman}, B., {Lawrence}, S., {Pun}, C.~S.~J.,
  {Garnavich}, P., {Kirshner}, R., {Crotts}, A., {Fransson}, C., {Li}, W.,
  {Panagia}, N., {Phillips}, M., {Schmidt}, B., {Sonneborn}, G., {Suntzeff},
  N., {Wang}, L., \& {Wheeler}, J.~C. 2003, \apj, 593, 809

\bibitem[{{Nussbaumer} \& {Schmutz}(1984)}]{nussbaumer84}
{Nussbaumer}, H. \& {Schmutz}, W. 1984, \aap, 138, 495

\bibitem[{{Osterbrock} \& {Ferland}(2006)}]{osterbrock06}
{Osterbrock}, D.~E. \& {Ferland}, G.~J. 2006, {Astrophysics of gaseous nebulae
  and active galactic nuclei}

\bibitem[{{Osterman} {et~al.}(2011){Osterman}, {Green}, {Froning},
  {B{\'e}land}, {Burgh}, {France}, {Penton}, {Delker}, {Ebbets}, {Sahnow},
  {Bacinski}, {Kimble}, {Andrews}, {Wilkinson}, {McPhate}, {Siegmund}, {Ake},
  {Aloisi}, {Biagetti}, {Diaz}, {Dixon}, {Friedman}, {Ghavamian}, {Goudfrooij},
  {Hartig}, {Keyes}, {Lennon}, {Massa}, {Niemi}, {Oliveira}, {Osten},
  {Proffitt}, {Smith}, \& {Soderblom}}]{osterman11}
{Osterman}, S., {Green}, J., {Froning}, C., {B{\'e}land}, S., {Burgh}, E.,
  {France}, K., {Penton}, S., {Delker}, T., {Ebbets}, D., {Sahnow}, D.,
  {Bacinski}, J., {Kimble}, R., {Andrews}, J., {Wilkinson}, E., {McPhate}, J.,
  {Siegmund}, O., {Ake}, T., {Aloisi}, A., {Biagetti}, C., {Diaz}, R., {Dixon},
  W., {Friedman}, S., {Ghavamian}, P., {Goudfrooij}, P., {Hartig}, G., {Keyes},
  C., {Lennon}, D., {Massa}, D., {Niemi}, S., {Oliveira}, C., {Osten}, R.,
  {Proffitt}, C., {Smith}, T., \& {Soderblom}, D. 2011, \apss, 157

\bibitem[{{Park} {et~al.}(2011){Park}, {Zhekov}, {Burrows}, {Racusin}, {Dewey},
  \& {McCray}}]{park11}
{Park}, S., {Zhekov}, S.~A., {Burrows}, D.~N., {Racusin}, J.~L., {Dewey}, D.,
  \& {McCray}, R. 2011, \apjl, 733, L35+

\bibitem[{{Pellerin} {et~al.}(2002){Pellerin}, {Fullerton}, {Robert}, {Howk},
  {Hutchings}, {Walborn}, {Bianchi}, {Crowther}, \& {Sonneborn}}]{pellerin02}
{Pellerin}, A., {Fullerton}, A.~W., {Robert}, C., {Howk}, J.~C., {Hutchings},
  J.~B., {Walborn}, N.~R., {Bianchi}, L., {Crowther}, P.~A., \& {Sonneborn}, G.
  2002, \apjs, 143, 159

\bibitem[{{Racusin} {et~al.}(2009){Racusin}, {Park}, {Zhekov}, {Burrows},
  {Garmire}, \& {McCray}}]{racusin09}
{Racusin}, J.~L., {Park}, S., {Zhekov}, S., {Burrows}, D.~N., {Garmire}, G.~P.,
  \& {McCray}, R. 2009, \apj, 703, 1752

\bibitem[{{Scuderi} {et~al.}(1996){Scuderi}, {Panagia}, {Gilmozzi}, {Challis},
  \& {Kirshner}}]{scuderi96}
{Scuderi}, S., {Panagia}, N., {Gilmozzi}, R., {Challis}, P.~M., \& {Kirshner},
  R.~P. 1996, \apj, 465, 956

\bibitem[{{Seaton}(1955)}]{seaton55}
{Seaton}, M.~J. 1955, Proceedings of the Physical Society A, 68, 457

\bibitem[{{Shigeyama} \& {Nomoto}(1990)}]{shigeyama90}
{Shigeyama}, T. \& {Nomoto}, K. 1990, \apj, 360, 242

\bibitem[{{Shull} {et~al.}(2010){Shull}, {France}, {Danforth}, {Smith}, \&
  {Tumlinson}}]{shull10}
{Shull}, J.~M., {France}, K., {Danforth}, C.~W., {Smith}, B., \& {Tumlinson},
  J. 2010, \apj, 722, 1312

\bibitem[{{Smith} {et~al.}(2005){Smith}, {Zhekov}, {Heng}, {McCray}, {Morse},
  \& {Gladders}}]{smith05}
{Smith}, N., {Zhekov}, S.~A., {Heng}, K., {McCray}, R., {Morse}, J.~A., \&
  {Gladders}, M. 2005, \apjl, 635, L41

\bibitem[{{Sonneborn} {et~al.}(1998){Sonneborn}, {Pun}, {Kimble}, {Gull},
  {Lundqvist}, {McCray}, {Plait}, {Boggess}, {Bowers}, {Danks}, {Grady},
  {Heap}, {Kraemer}, {Lindler}, {Loiacono}, {Maran}, {Moos}, \&
  {Woodgate}}]{sonneborn98}
{Sonneborn}, G., {Pun}, C.~S.~J., {Kimble}, R.~A., {Gull}, T.~R., {Lundqvist},
  P., {McCray}, R., {Plait}, P., {Boggess}, A., {Bowers}, C.~W., {Danks},
  A.~C., {Grady}, J., {Heap}, S.~R., {Kraemer}, S., {Lindler}, D., {Loiacono},
  J., {Maran}, S.~P., {Moos}, H.~W., \& {Woodgate}, B.~E. 1998, \apjl, 492,
  L139+

\bibitem[{{Spitzer}(1978)}]{spitzer78}
{Spitzer}, L. 1978, {Physical processes in the interstellar medium}

\bibitem[{{van Adelsberg} {et~al.}(2008){van Adelsberg}, {Heng}, {McCray}, \&
  {Raymond}}]{adelsberg08}
{van Adelsberg}, M., {Heng}, K., {McCray}, R., \& {Raymond}, J.~C. 2008, \apj,
  689, 1089

\bibitem[{{Walborn} {et~al.}(1993){Walborn}, {Phillips}, {Walker}, \&
  {Elias}}]{walborn93}
{Walborn}, N.~R., {Phillips}, M.~M., {Walker}, A.~R., \& {Elias}, J.~H. 1993,
  \pasp, 105, 1240

\bibitem[{{Walker} \& {Suntzeff}(1990)}]{walker90}
{Walker}, A.~R. \& {Suntzeff}, N.~B. 1990, \pasp, 102, 131

\bibitem[{{Xu} {et~al.}(1992){Xu}, {McCray}, {Oliva}, \& {Randich}}]{xu92}
{Xu}, Y., {McCray}, R., {Oliva}, E., \& {Randich}, S. 1992, \apj, 386, 181

\bibitem[{{Zhekov} {et~al.}(2006){Zhekov}, {McCray}, {Borkowski}, {Burrows}, \&
  {Park}}]{zhekov06}
{Zhekov}, S.~A., {McCray}, R., {Borkowski}, K.~J., {Burrows}, D.~N., \& {Park},
  S. 2006, \apj, 645, 293

\end{thebibliography}

\end{document}